\documentclass[sigconf]{sty/acmart}

\newif\ifarxiv \arxivtrue

\setcopyright{none}
\renewcommand\footnotetextcopyrightpermission[1]{%
  \footnotetext{\textbf{Accepted at ACM CCS 2026.} This is the extended version
  of a paper appearing in the Proceedings of the 2026 ACM SIGSAC Conference on
  Computer and Communications Security (CCS '26), November 15--19, 2026, The
  Hague, Netherlands. It additionally contains the full technical appendix.
  Please cite the published version:
  https://doi.org/10.1145/3830454.3846550}}

\usepackage{graphicx}
\usepackage{xcolor}
\newcommand{\nanzi}[1]{}
\newcommand{\bowen}[1]{}
\newcommand{\weiheng}[1]{}
\newcommand{\kangjie}[1]{}

\usepackage{colortbl}

\usepackage{silence}
\usepackage{listings}
\usepackage{xcolor}
\usepackage{multirow}
\usepackage{algorithm}
\usepackage{algorithmicx}
\usepackage{algpseudocode}
\usepackage{graphicx}
\usepackage{subcaption}

\lstdefinelanguage{Solidity}{
  keywords={
    pragma, solidity, import, contract, interface, function, returns, return,
    public, external, internal, private, view, pure, payable,
    address, uint, uint256, bytes, bytes32, bool, mapping, memory, calldata, storage,
    if, else, for, while, do, break, continue,
    try, catch, revert, require, assert,
    emit, event, modifier, constant, immutable, new
  },
  keywordstyle=\color{blue}\bfseries,
  ndkeywords={true,false,this,super,msg,block,tx},
  ndkeywordstyle=\color{teal}\bfseries,
  sensitive=true,
  comment=[l]{//},
  morecomment=[s]{/*}{*/},
  commentstyle=\color{gray}\itshape,
  stringstyle=\color{purple},
  morestring=[b]",
}

\lstdefinestyle{quote}{
  basicstyle=\ttfamily\scriptsize,
  columns=fullflexible,
  keepspaces=true,
  showstringspaces=false,
  breaklines=true,
  breakatwhitespace=true,
  frame=none,
  numbers=none,
  xleftmargin=0.4em,
  xrightmargin=0.2em,
  aboveskip=0.3em,
  belowskip=0.3em,
  tabsize=2
}

\lstdefinestyle{code}{
  basicstyle=\ttfamily\scriptsize,
  columns=fullflexible,
  keepspaces=true,
  showstringspaces=false,
  breaklines=true,
  breakatwhitespace=true,
  frame=none,
  numbers=none,
  xleftmargin=0.4em,
  xrightmargin=0.2em,
  aboveskip=0.3em,
  belowskip=0.3em,
  tabsize=2,
  language=C,
  keywordstyle=\color{blue}\bfseries,
  commentstyle=\color{gray}\itshape,
  stringstyle=\color{purple},
  emph={targets,calldatas,transfer,changeProxyAdmin},
  emphstyle=\bfseries
}

\lstdefinestyle{simcode}{
  basicstyle=\ttfamily\scriptsize,
  columns=fullflexible,
  keepspaces=true,
  showstringspaces=false,
  breaklines=true,
  breakatwhitespace=true,
  frame=lines,
  framerule=0.4pt,
  framesep=4pt,
  backgroundcolor=\color{black!3},
  numbers=none,
  xleftmargin=0pt,
  xrightmargin=0pt,
  aboveskip=0.4em,
  belowskip=0.4em,
  tabsize=2,
  language=C,
  keywordstyle=\color{blue}\bfseries,
  commentstyle=\color{gray}\itshape,
  stringstyle=\color{purple},
  emph={warp,executeVariants,prank,queueVariants,rollAndWarp},
  emphstyle=\color{teal}\bfseries
}

\lstdefinestyle{prompttpl}{
  basicstyle=\ttfamily\scriptsize,
  columns=fullflexible,
  keepspaces=true,
  showstringspaces=false,
  breaklines=true,
  breakatwhitespace=true,
  frame=lines,
  framerule=0.4pt,
  framesep=4pt,
  backgroundcolor=\color{black!3},
  numbers=none,
  xleftmargin=0pt,
  xrightmargin=0pt,
  aboveskip=0.4em,
  belowskip=0.4em,
  tabsize=2,
  emph={Knowledge,Detection,Task,System,Prompt,User},
  emphstyle=\color{teal}\bfseries
}

\lstdefinestyle{prompt}{
  basicstyle=\ttfamily\scriptsize,
  columns=fullflexible,
  keepspaces=true,
  showstringspaces=false,
  breaklines=true,
  breakatwhitespace=true,
  frame=single,
  rulecolor=\color{black!15},
  framerule=0.3pt,
  xleftmargin=0.0em,
  xrightmargin=0.0em,
  aboveskip=0.4em,
  belowskip=0.4em,
  tabsize=2
}

\usepackage{pifont}
\usepackage{tabularx}
\hypersetup{citecolor=teal,linkcolor=violet}
\usepackage{amsopn}
\usepackage{xspace,fancyvrb}
\usepackage{booktabs}
\usepackage{array}

\usepackage{siunitx}

\usepackage{balance}

\usepackage{comment}
\usepackage{amsthm}
\usepackage{mdframed}

\newtheorem{definition}{Definition} 

\usepackage{tikz}
\newcommand*{\circled}[1]{\lower.7ex\hbox{\tikz\draw (0pt, 0pt)%
    circle (.5em) node {\makebox[1em][c]{\small #1}};}}

\newcommand{\KL}[1]{}
\newcommand{\WB}[1]{}
\newcommand{\TODO}[1]{}
\newcommand{\BW}[1]{}

\fvset{fontsize=\scriptsize,xleftmargin=8pt,numbers=left,numbersep=5pt}

\input{fmt}

\def\Snospace~{\S{}}

\algrenewcommand{\algorithmiccomment}[1]{\hfill\textcolor{Sepia}{\(\triangleright\) #1}}

\newif\ifdraft\drafttrue
\newif\ifnotes\notestrue
\ifdraft\else\notesfalse\fi

\input{glyphtounicode}
\newcolumntype{R}[1]{>{\raggedleft\let\newline\\\arraybackslash\hspace{0pt}}p{#1}}

\newcommand{\squishlist}{
\begin{itemize}[noitemsep,nolistsep]
  \setlength{\itemsep}{-0pt}
}
\newcommand{\squishend}{
  \end{itemize}
}

\usepackage{tikz}

\usepackage{xstring}
\newcommand{\PP}[1]{
\noindent{\bf \IfEndWith{#1}{.}{#1}{#1.}}
}

\newcommand{\boxbeg}{
\vspace{2px}
\noindent\begin{tabular}{|l|}\hline
\begin{minipage}{3.2in}
\vspace{2px}
\noindent
}

\newcommand{\boxend}{
\vspace{2px}
\end{minipage}\\ \hline
\end{tabular}
\vspace{-10pt}
}

\copyrightyear{2026}
\acmYear{2026}
\setcopyright{cc}
\setcctype{by}
\acmConference[CCS '26]{Proceedings of the 2026 ACM SIGSAC Conference on Computer and Communications Security}{November 15--19, 2026}{The Hague, Netherlands}
\acmBooktitle{Proceedings of the 2026 ACM SIGSAC Conference on Computer and Communications Security (CCS '26), November 15--19, 2026, The Hague, Netherlands}
\acmDOI{10.1145/3830454.3846550}
\acmISBN{979-8-4007-2871-6/2026/11}

\title{Mind the Gap: Detecting Description-Execution Mismatch Attacks in DAO Governance}

\author{Bowen Cai}
\affiliation{%
  \institution{University of Minnesota}
  \city{Minneapolis}
  \state{Minnesota}
  \country{USA}}
\email{cai00254@umn.edu}

\author{Nanzi Yang}
\affiliation{%
  \institution{Old Dominion University}
  \city{Norfolk}
  \state{Virginia}
  \country{USA}}
\email{nzyang@stu.xidian.edu.cn}

\author{Weiheng Bai}
\affiliation{%
  \institution{University of Minnesota}
  \city{Minneapolis}
  \state{Minnesota}
  \country{USA}}
\email{bai00093@umn.edu}

\author{Youshui Lu}
\affiliation{%
  \institution{Xi'an Jiaotong University}
  \city{Xi'an}
  \country{China}}
\email{yolu6176@uni.sydney.edu.au}

\author{Yajin Zhou}
\affiliation{%
  \institution{The Chinese University of Hong Kong}
  \city{Hong Kong}
  \country{China}}
\email{yajin@yajin.org}

\author{Kangjie Lu}
\affiliation{%
  \institution{University of Minnesota}
  \city{Minneapolis}
  \state{Minnesota}
  \country{USA}}
\email{kjlu@umn.edu}

\renewcommand{\shortauthors}{Cai et al.}



\ifdefined\DRAFT
\fi

\begin{document}

\begin{abstract}
Decentralized autonomous organizations (DAOs) make protocol changes through the proposal-based process: (1) Initiators submit a proposal; (2) DAO members vote for it based on the proposal description, and if it passes, (3) the project executes the code behind the proposal. Such a process is inherently vulnerable to \emph{deceptive proposals}: the description intent and the actual code execution may mismatch. A malicious proposer could submit a proposal with a benign-looking description to pass voting, while the executed code performs harmful actions, such as transferring funds or taking control of the protocol, which we call \emph{Description-Execution Mismatch (DEMI) attack}.

In this paper, we present the first systematic framework for DEMI detection in real DAO governance. First, the diversity of DAO deployments makes it difficult to design a unified analysis that
can apply and scale to different proposals. 
To address this challenge, we propose a \emph{DAO-agnostic simulation framework}; its core first builds a per-DAO governance profile from historical on-chain transactions, then performs live proposal simulation by driving each new proposal through the full governance lifecycle to obtain its execution behaviors.
Second, the incompatibility between free-form descriptions and structured execution traces makes consistency checking non-trivial. We address this via an \emph{evidence-mapping} paradigm that requires an LLM to locate explicit per-action textual justifications, rather than issuing a holistic consistency judgment, substantially improving precision and recall over direct querying.

We evaluate our system on a large-scale dataset of real-world Ethereum DAO governance activity. Our lifecycle simulation successfully derives execution results for 92.7\% of active DAOs and 89.3\% of executed proposals, substantially exceeding the coverage of existing governance platforms. Our DEMI detector achieves a mean precision of 81.7\% and a mean recall of 98.3\% under stratified cross-validation, and its underlying evidence-mapping design generalizes across LLM vendors rather than depending on any single model. Under a systematic red-team/blue-team evaluation, the Robustness Guard defends the large majority of adaptive attacks even against an adversary that knows the detector, and this robustness appears largely structural: it generalizes to held-out proposals, and in our evaluation an automated blue team armed with the same strong model did not improve on it. Together, our results demonstrate that verifiable and robust proposal mismatch checking is both feasible and practical at scale, providing a critical security primitive for DAO governance.

\end{abstract}

\maketitle

\sloppy

\section{Introduction}

Decentralized autonomous organizations (DAOs) increasingly rely on on-chain governance to coordinate critical protocol decisions, including treasury management, parameter updates, and access control changes. Governance proposals are no longer a peripheral coordination mechanism; they directly determine how assets move and how control is exercised in deployed systems~\cite{DAO_process_explain_arxiv, Good_SoK_for_DAO, Security_for_DeFi_tokenomics}. As a result, governance itself has become a security-critical attack surface, with recent incidents demonstrating that a single malicious proposal can cause irreversible financial losses~\cite{Security_for_DeFi_tokenomics, DAO_security_DeFiWorkshop, DAO_Governance_security_review_tosem}.

In a typical DAO governance process, proposal acceptance follows a three-phase lifecycle: \texttt{propose} $\rightarrow$ \texttt{vote}
$\rightarrow$ \texttt{execute}. During this lifecycle, community members evaluate a proposal \emph{ex ante} based on a human-facing natural-language description presented before voting, while the system ultimately enforces the proposal \emph{ex post} through an encoded execution payload that deterministically produces on-chain execution traces. As a result, whether a governance decision is carried out as intended depends critically on the alignment between what voters are shown before approval and what the code actually executes after approval.

The reason DAO governance adopts this vote-before-execution workflow lies fundamentally in the design philosophy of DAOs: the decentralization of rights~\cite{DAO_Governance_security_review_tosem}.  By replacing centralized administrators with community-driven decision-making, DAOs intentionally eliminate privileged authorities and rely on community voting to authorize proposals. As a result, proposals must be evaluated and approved by the community execution \emph{ex ante}. This design shifts both decision power and responsibility to the community members.

However, most community members lack the technical expertise required to reason about smart contract execution, governance-specific call patterns, or subtle control-flow implications~\cite{Good_SoK_for_DAO}. They are structurally constrained to rely on descriptions that abstract away execution details, while the governance process itself provides no binding mechanism to enforce alignment between what voters see and what the system ultimately executes.

Such a process inevitably leads to a critical security
problem, namely \emph{Description--Execution Mismatch (DEMI)}, where there is a mismatch between the proposal description that informs community voting
(\emph{what is claimed}) and the execution results observed
on-chain (\emph{what is enforced}). 
A malicious community member can exploit this mismatch to steal funds, or take control of the entire DAO~\cite{DAO_Governance_security_review_tosem}, which we refer to as the \emph{DEMI attack}. More specifically, during proposal submission, an attacker can craft a benign-looking description that appears routine or socially legitimate, thereby misleading the community during the voting phase. After the proposal is approved, the attacker's malicious execution payload (e.g., fund-draining transactions or privilege-escalation calls) is executed automatically on-chain, and the resulting damage is often irreversible.
This makes proposal mismatch a particularly dangerous class of governance
vulnerability: it converts the collective trust into a malicious action at execution time.

Although the DEMI attacks are practical and lead to severe consequences, existing techniques can not systematically identify such risks. More specifically, existing governance tools and platforms~\cite{Arbitrum_Governance,Compound_Governance,daostack,Uniswap_Governance,Agora,Tally_explore,Snapshot,DAO_Governance_security_review_tosem} are typically designed for \emph{ex-post attack monitoring}, rather than \emph{ex-ante prevention} through enforceable execution simulation and verification. They mainly surface proposal artifacts, such as natural-language descriptions, calldata, and execution payloads, but largely stop short of deriving a deterministic and verifiable execution trace under real execution contexts. Consequently, malicious results often remain invisible until execution, at which point governance actions become difficult or impossible to reverse, leading to potentially irreversible losses.

In this paper, we propose the first systematic framework for \emph{ex ante} DEMI detection: one that operates after a proposal is submitted but before it is executed, so the community can still contest or reverse a malicious proposal. Our key observation is that mismatch detection is a conformance checking between two heterogeneous representations. On one hand, the governance proposal is presented to the community in natural language description before voting. On the other hand, the proposal is to be enforced on-chain as a set of execution traces of protocol operations after voting. As a result, detecting mismatches reduces to \emph{cross-checking} two representations of the same decision: (i) the semantic intent conveyed in the description, and (ii) the effective execution semantics induced by the proposal payload. Therefore, a practical mismatch detector must scale to large numbers of diverse proposals with heterogeneous DAOs, and remain robust against evasive or strategically crafted descriptions that aim to mask mismatch behaviors.

The first challenge is scalability across diverse proposal simulations. In practice, governance mechanisms vary significantly across DAOs, including their proposal submission methods, voter compositions, execution payload structures, and so on. All these diverse scenarios construct the proposal process unique to a specific DAO. As a result, we need to develop a unified method for simulating diverse processes across different DAO scenarios.

Second, even with the execution traces, there still exists incompatibility between descriptions and underlying execution traces. On one hand, the execution trace is structured, which consists of a sequence of function calls. Its naming styles and calling methods vary significantly across different DAOs. On the other hand, descriptions are free-form natural language, which express high-level intent rather than concrete execution steps, and the same intent may correspond to different function names and execution methods.

To solve the first challenge, we propose a \emph{lifecycle-based simulation} technique. Our key observation is that governance proposal histories are publicly observable on-chain. Although different DAOs use different functions, proposals within the same DAO reuse a consistent set of lifecycle-specific functions across the propose, vote, and execution stages. As a result, we first build a per-DAO \emph{governance profile} by extracting these stage-specific functions and governance parameters from historical transactions. For each newly submitted proposal, we then assemble its simulation inputs from the profile and perform \emph{live proposal simulation}: driving the proposal through the full governance lifecycle to obtain a verifiable execution trace. Crucially, all required inputs are available at proposal submission time, which is why our simulation can run during the voting phase rather than requiring \emph{ex post} data.

To solve the second challenge, we propose a \emph{semantic-based DEMI guard} built on Large Language Model (LLM). Proposal descriptions are authored by independent developers across hundreds of DAOs with highly heterogeneous formats, vocabularies, and writing conventions. This heterogeneity makes rule-based keyword matching and traditional NLP methods impractical due to prohibitive false-positive noise. While a direct LLM query already substantially outperforms such approaches through zero-shot semantic reasoning, naively asking the LLM for a holistic consistency judgment still misses a large fraction of mismatches. Our key design insight is therefore to reframe detection as an \emph{evidence-mapping} problem: rather than requesting an overall verdict, we require the LLM to locate an explicit supporting span in the description for each executed action individually. This evidence-seeking paradigm mirrors human auditing practice, grounds LLM reasoning in concrete textual evidence, and substantially improves both recall and precision over direct querying.

Besides these two challenges, a common concern of our detection is robustness. The root cause of a DEMI attack is the semantic deception between what is claimed in the description and what is enforced in execution traces. As a result, attackers can deliberately introduce vague or over-permissive phrasing that misleads the DEMI guard proposed by us. 

To solve this problem, we propose a \emph{robustness guard} technique. Rather than merely checking whether some description span can be matched to an action, the robustness guard evaluates whether the evidence is grounded, sufficiently constraining, and aligned with the proposal's stated purpose. By rejecting fabricated, over-permissive, or purpose-misaligned justifications, this guard prevents adversarial ambiguity from undermining automated mismatch verification.

Because our detector's prompt and risk taxonomy are published, robustness must be demonstrated against an attacker who knows them, not merely asserted. We therefore treat adversarial evaluation as a distinct component of this work: a systematic \emph{red-team/blue-team} study in which an informed adversary iteratively crafts evasive descriptions while an automated blue team hardens the guard, plus a prompt-injection stress test. This measures what an informed attacker can actually achieve rather than reporting a single static number.

To evaluate our system, we run it on a large-scale dataset of real-world DAOs. Our lifecycle-based simulation derives execution traces for 92.7\% of DAOs and 89.3\% of proposals. The DEMI detector achieves a mean precision of 81.7\% and a mean recall of 98.3\% under stratified 4-fold cross-validation. This accuracy is not an artifact of incidental choices: an ablation shows it already holds with a single reasoning pass rather than multi-sample voting, and the evidence-mapping advantage holds across eight LLMs spanning three vendors, sustaining 83--99\% recall on every one. A systematic red-team/blue-team study further shows that, in our tested setting, the guard defends about 80\% of a realistic gray-box adversary's adaptive attacks, and the majority even under a worst-case white-box adversary that holds our exact detector. Strikingly, in our study a strong automated blue team with full knowledge of the detector did not improve on this base configuration, suggesting that the robustness stems largely from the evidence-mapping structure rather than from a set of patchable rules. These results demonstrate that scalable, accurate, and robust proposal mismatch checking is feasible.

Furthermore, our tool finds a set of severe vulnerabilities with real security impacts. We identified 69 proposals exhibiting clear malicious intent, including 9 that have already been officially confirmed by the project parties. In addition, 475 proposals reflect widespread irresponsible governance practices that undermine transparency and informed decision-making. These findings show that proposal mismatch is both practically exploitable and pervasive in real-world DAO governance.

In summary, this paper makes the following contributions:

\noindent\textbf{Novel attack surface:}
While prior work has documented individual DAO governance attacks and description--execution discrepancies~\cite{DAO_Governance_security_review_tosem, DAO_security_DeFiWorkshop}, we provide the first systematic formalization and large-scale measurement of \emph{DEMI} as an attack surface in DAO governance, where mismatches between proposal descriptions and execution traces can be exploited to mislead community members and enable fund drainage or governance control escalation.

\noindent\textbf{Pre-execution detection framework:}
We introduce a methodology for detecting DEMI before a proposal is executed and while governance decisions can still be contested. By simulating the full \texttt{propose-vote-queue-execute} lifecycle using only public on-chain data, our approach derives authentic execution traces scalably across heterogeneous DAOs. On top of these traces, an evidence-mapping detector performs accurate mismatch checking that generalizes across LLM vendors rather than depending on any single model, and a robustness guard keeps detection robust against adversarially crafted descriptions. A systematic red-team/blue-team study indicates this robustness is largely structural: it withstands adaptive white-box attacks, generalizes to unseen proposals, and, in our study, an automated blue team with full knowledge of the detector did not improve on the base design, suggesting the heuristic already captures much of what the detection task requires.

\noindent\textbf{Real-world impact:}
We conduct the first large-scale empirical study of proposal mismatch across
real-world DAO deployments, analyzing over 10,000 executed proposals.
Our analysis uncovers 69 previously unreported malicious proposals that expose
DAO treasuries and governance control to significant risk, collectively involving
over 500 ETH, as well as 475 additional irresponsible
proposals that reflect widespread governance hygiene issues.

\section{Background}

\subsection{DAO Governance Proposal}
\begin{figure}[t]
    \centering
    \includegraphics[width=0.95\linewidth]{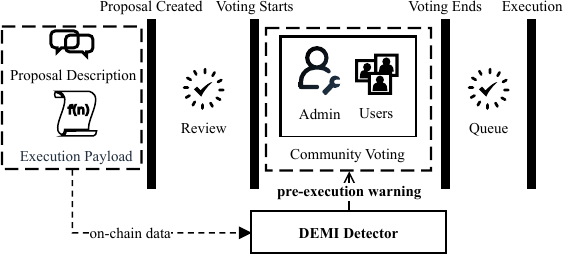}
    \caption{The Overview of DAO Governance}
    \label{fig:governance-overview}
\end{figure}
Decentralized Autonomous Organizations (DAOs) rely on on-chain governance to coordinate collective decision-making without centralized control~\cite{Good_SoK_for_DAO}. In most DAOs, governance is implemented through smart contracts that define how proposals are submitted, voted on, queued(optional), and executed. Although execution is ultimately enforced by code, the decision-making process remains human-driven: voters must interpret proposal information to infer intent and anticipate consequences.

A typical governance proposal follows a multi-stage lifecycle, as shown in \autoref{fig:governance-overview}. It begins when a proposer submits a proposal that bundles two complementary artifacts. The first is the \emph{proposal description}, a natural-language, human-facing statement of intent and rationale. It is the primary information source for community discussion and voting, yet it is not machine-enforced and therefore cannot directly constrain what will happen on-chain. The second is the \emph{execution payload}, which specifies the concrete on-chain actions to be carried out if the proposal is approved, typically encoded as calldata and function calls to one or more smart contracts. Unlike the description, the payload is what the governance system will actually execute, but its implications are unclear without simulation. Not every governance action carries such a payload: some DAOs also use purely social or signaling proposals that trigger no on-chain execution, or delegate follow-up actions to a foundation. These are outside our scope by construction, since DEMI concerns a mismatch between a description and an on-chain execution; our study targets the proposals that do carry an executable payload.

After submission, the proposal enters the voting phase, during which token holders evaluate the proposal, often relying heavily on its description. If approved, many DAOs impose a timelock period that provides a final window to inspect the queued payload before it can be executed. Once executed on-chain, the proposal produces an \emph{execution trace}: a runtime record of the realized behavior, including the sequence of contract calls and state transitions. The trace captures the proposal's concrete effects and is not fully predictable from the static payload alone due to dynamic control flow and external dependencies, for example a transfer amount computed from an on-chain oracle price at execution time, or a call whose effect depends on the runtime state of another contract.

However, existing governance workflows largely assume that the proposal descriptions are aligned with real execution, and there is no enforced mechanism to verify that the description faithfully matches the proposal's actual on-chain behavior, motivating our focus on \emph{description--execution mismatch}.

\subsection{Description--Execution Mismatch}

Description--execution mismatch (DEMI) asks whether a \emph{proposal's description} remains faithful to what the proposal will actually do on-chain. Concretely, once a proposal is approved, its \emph{execution payload} is invoked to perform a set of contract calls; the resulting \emph{execution trace} reveals the realized behavior and effects. A proposal is matched if the execution trace matches what a reasonable community member would infer from the description. Otherwise, there exists a mismatch between the proposal description and execution.

However, there exists a gap between voting and real execution in today's governance workflow. During deliberation and voting, the community primarily observes and reasons about the description, while the payload's semantics are encoded and often difficult to interpret without specialized tooling. The trace, the most direct representation of true effects, only exists after execution, when it is already too late to prevent damage. This separation creates room for exploitation: a proposal can present a benign description while embedding a payload whose executed trace produces materially different outcomes (e.g., hidden privilege changes, asset transfers, or parameter updates with unintended side effects). This motivates us to formalize and measure the description--execution mismatch, and use a real example to show the attacks are real with severe consequences (see more details in \autoref{sec:motivation}).


\section{Motivation}\label{sec:motivation}
In this section, we first present our threat model and assumptions. We then formally define when a description--execution mismatch (DEMI) occurs and characterize it as a new class of governance attacks. Finally, we use a real attack to illustrate the details.

\subsection{Threat Model}
Our central assumption is that voters form their approval decisions primarily from a proposal's natural-language description and UI-rendered summaries, rather than by decoding its raw execution payload. This reflects practical governance workflows: modern DAOs increasingly rely on standardized governance tooling (e.g., proposal forums, off-chain discussion, and on-chain executors), and in practice only a small fraction of participants manually decode raw execution payloads or reason about low-level call traces, due to limited time, expertise, and the complexity of multi-call transactions. Prior governance participation suggests that such asymmetric scrutiny is common in real-world voting behavior~\cite{Good_SoK_for_DAO, DAO_Whale_colusion}. The semantics conveyed by proposal descriptions therefore often dominate voters' understanding and decision-making.

Based on these real-world governance practices, we consider the following adversary model. On one hand, we assume the underlying blockchain, smart contract execution, and governance infrastructure operate correctly (i.e., no protocol-level compromise), and we do not rely on low-level smart contract bugs to mount the attack. On the other hand, the attacker's goal is to mislead voters into approving proposals whose executed effects deviate from what voters reasonably infer from the description, typically to steal funds or seize control of the DAO. The adversary's resources are constrained by the governance system itself. The adversary cannot bypass established safeguards such as voting thresholds, quorum requirements, timelocks, or the prescribed propose-vote-queue-execute lifecycle. The attacker operates under the same governance contracts as everyone else and can only manipulate the proposal's content (an arbitrary \emph{proposal description} and \emph{execution payload}) subject to the same submission rules and on-chain enforcement as benign proposers.

In the following, we show that even under this restricted adversary model, an attacker can still launch description--execution mismatch attacks with severe consequences.

\subsection{Threat Definition}

Existing governance platforms provide no support for systematically assessing whether a proposal's described intent matches its actual on-chain effects, motivating the need for a formal notion of description--execution mismatch (DEMI). In practice, these platforms aim to improve transparency by exposing proposal-related information to voters, typically including the human-facing description and, in some cases, the encoded execution payload. However, execution traces are often unavailable, and when execution previews are shown, they are proposer-supplied and lack independent provenance guarantees.

Based on these observations, we introduce a formal definition of governance proposals. We view a governance proposal through three representations: the human-facing \emph{proposal description} $d \in \mathcal{D}$, the encoded \emph{execution payload} $p \in \mathcal{P}$ (e.g., calldata specifying on-chain actions), and the resulting \emph{execution trace} $\tau \in \mathcal{T}$ that captures runtime behavior upon execution. The payload $p$ is what the governance contract actually executes, but voters reason over $d$; the trace $\tau=\psi(p)$ is the objective bridge between them. To make textual intent comparable to runtime behavior, we introduce a mapping $\phi: \mathcal{D} \rightarrow \mathcal{T}$ that maps a natural-language description into the trace-level behavior a voter would \emph{expect} it to produce, and a predicate $\mathrm{Cons}: \mathcal{T} \times \mathcal{T} \rightarrow \{\mathit{True}, \mathit{False}\}$ that holds when an expected behavior is entailed by (consistent with) an actual trace.

\begin{definition}[Execution Trace Derivation]\label{def:simulation}
Given a payload $p \in \mathcal{P}$, an execution trace $\tau \in \mathcal{T}$ is \emph{derivable} if it is produced by an execution derivation procedure: $\tau = \psi(p)$, such that any third party can reproduce or validate $\tau$ from $p$ under the same on-chain state assumptions.
\end{definition}

\begin{definition}[Description--Execution Mismatch, DEMI]\label{def:inconsistency}
Let $d \in \mathcal{D}$ be a proposal description and $p \in \mathcal{P}$ be its
execution payload. Writing the derived trace explicitly as $\psi(p)$, a
\emph{DEMI} occurs when $\mathrm{Cons}\big(\phi(d),\ \psi(p)\big)=\mathit{False}$,
i.e., the behavior a voter would expect from the description $\phi(d)$ is not
consistent with the behavior the payload actually produces $\psi(p)$.
\end{definition}

In other words, DEMI arises when the human-facing description
used for voting (ex ante) fails to faithfully reflect the on-chain behavior
enforced after execution (ex post), under an authentic execution trace.

Under our formulation, verifiable DEMI requires: (i) deriving the execution trace via $\psi(p)$ and (ii) checking alignment automatically via $\mathrm{Cons}(\phi(d),\psi(p))$, addressing the lack of independent execution results and the reliance on human-centered verification in existing tooling.

\subsection{Real World Example: DEMI Attack on Pepe Cash Governance}

\begin{figure}[t]
    \centering

    \begin{subfigure}{\linewidth}
        \begin{lstlisting}[style=quote]
"After initial preliminary voting within the community discussions on Discord, it was proposed to redistribute the $PCASH tokens in the 'Claim Safe' address 0x6232xxx. The consensus among voters was that this proposal should be approved on-chain to advise developers to complete the process formally. It was also agreed that 66% (two-thirds) of the wallet holdings should be transferred to the marketing wallet 0x9c44xxx, amounting to a total of 4,026,000,000,000 tokens."
        \end{lstlisting}
        \caption{Proposal Description}
        \label{fig:motivation-desc}
    \end{subfigure}

    \begin{subfigure}{\linewidth}
        \begin{lstlisting}[style=code]
execute(
  targets   = [0x4FAB...830e],
  values    = [0],
  calldatas = [0xa9059...000],
  proposer  = 0xe431...40e4
)
        \end{lstlisting}
        \caption{Execution Payload}
        \label{fig:motivation-payload}
    \end{subfigure}

    \begin{subfigure}{\linewidth}
        \begin{lstlisting}[style=code]
transfer(0xfd63...c15f, 1.26207e31)
        \end{lstlisting}
        \caption{Key Execution Trace}
        \label{fig:motivation-trace}
    \end{subfigure}

    \caption{\texttt{Proposal-779xx} in Pepe Cash DAO: benign description versus malicious execution.}
    \label{fig:motivation}
\end{figure}

\begin{figure}[t]
    \centering
    \includegraphics[width=\linewidth]{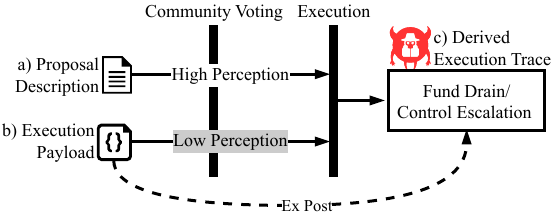}
    \caption{Diagram of the proposal attack.}
    \label{fig:attack_example}
\end{figure}

Viewed through our formal definitions (\autoref{def:simulation} and \autoref{def:inconsistency}), the core governance failure in DEMI attacks is not subtle: during the voting phase, the community lacks access to an \emph{execution trace} $\tau=\psi(p)$. As a consequence, voters cannot evaluate whether the proposal description $d$ and the actual execution behavior $\tau$ satisfy $\mathrm{Cons}(\phi(d),\tau)=1$. Governance decisions are therefore made under an information asymmetry: what voters see and reason about ex ante differs fundamentally from what the system will enforce ex post.

We illustrate this abstract gap with a real-world governance incident from the Pepe Cash DAO. In this project, Proposal-779~\cite{Tally_PEPE_CASH_779} was submitted by a community member and closely mirrored the description of a previously approved, legitimate proposal (Proposal-913~\cite{Tally_PEPE_CASH_913}). As shown in \autoref{fig:motivation}, the description of Proposal-779 described the action as a routine redistribution of {PCASH} tokens following prior community discussions, explicitly claiming the source wallet, the destination marketing wallet, and a precise transfer ratio. From a voter's perspective, this narrative strongly suggested a benign and familiar operational task.

However, this perception was formed entirely at the description level. As depicted in \autoref{fig:motivation-payload} and \autoref{fig:motivation-trace}, the encoded execution payload ultimately produced a very different execution trace:
\texttt{transfer(0xfd63...,\ 1.26e13)}. Rather than implementing the claimed redistribution, the proposal execution transferred a massive amount of tokens to an address controlled by the proposer, resulting in direct fund extraction. This divergence between the human-facing description and the derived execution trace constitutes a concrete instance of DEMI attack under our definition: although a description $d$ was available, voters had no execution trace $\tau=\psi(p)$ to verify whether $\mathrm{Cons}(\phi(d),\tau)=True$ held prior to execution.

\autoref{fig:attack_example} summarizes this attack vector from a systemic perspective. The proposal lifecycle exposes voters only to the description during the \texttt{propose} and \texttt{vote} stages. However, the result of \emph{execution payload} remains unknown until after execution. Once the proposal passes, the system enforces the encoded behavior automatically, converting misplaced trust into irreversible on-chain actions.

Importantly, the Pepe Cash incident is not an isolated anomaly. Since 2016, at least 33 DAO governance attacks have been documented~\cite{Good_SoK_for_DAO}. Our inspection shows that DEMI plays a central role in 26 of these cases, involving approximately \$201 million USD in losses. These observations underscore that proposal mismatch is a recurrent and economically significant security risk. Without systematic mismatch checking, governance communities remain structurally exposed to such attacks.

\section{DEMI Detection Framework}\label{sec:implementation}

\begin{figure*}[t]
    \centering
    \includegraphics[width=0.78\linewidth]{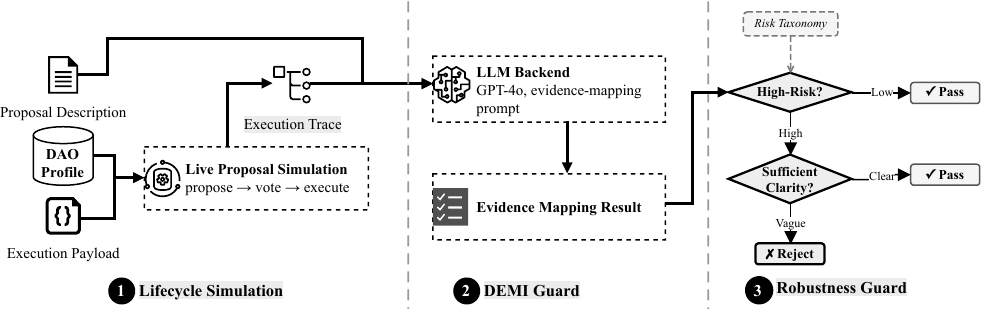}
    \caption{The Framework for DEMI Detector}
    \label{fig:framework}
\end{figure*}

In this section, we first introduce the limitations of existing tools. Subsequently, to systematically detect \emph{description--execution mismatch (DEMI) attacks}, we present our approach and provide details.

\subsection{Limitations of Existing Tools}\label{subsec:limitation}

Existing DAO governance tools fail to defend against DEMI attacks for two
structural reasons.
\emph{(1) Lack of independent and reproducible simulation.}
Simulating proposal execution typically requires privileged administrator roles
or state-dependent execution contexts that are not publicly available, so
execution previews, when offered at all, are proposer-supplied and cannot be
independently reproduced or validated by third parties. For example, the
simulated outcome Tally may display is submitted by the proposer, so a voter
cannot confirm that it corresponds to the payload that will actually execute.
\emph{(2) Human-centered verification bottleneck.}
Even when previews exist, verifying them is delegated to voters, many of whom, we
infer, do not audit execution traces or reconcile them against the description.
In the Pepe Cash attack (\autoref{sec:motivation}), for instance, the description
named a specific marketing wallet and amount, yet approving voters did not check
the calldata that ultimately moved the funds. This leaves the workflow
non-scalable and exposed to misleading or evasive proposals.

These limitations are reflected in widely used governance tools.
Snapshot~\cite{Snapshot} serves as an off-chain voting interface: it exposes
descriptions and, when provided, proposer-submitted payloads, but offers no
mechanism to derive execution results, fully offloading mismatch reasoning to
voters. Tally~\cite{Tally_explore} extends this model with on-chain governance
and proposal statistics, but its simulated results are proposer-submitted, as
noted above, leaving the same trust and scalability issues unresolved. Together,
these observations motivate an automated and verifiable approach to DEMI
detection, which we introduce next.

\subsection{Challenges and Overview}
\label{subsec:challenges-overview}

Our framework is motivated by the observation that proposal mismatch detection fundamentally reduces to a \emph{conformance checking} problem between two heterogeneous representations of the same governance decision: the natural-language description used for voting, and the execution trace induced by the on-chain payload. Practical detection therefore requires systematic \emph{cross-checking} between high-level semantic intent and low-level execution semantics.

This formulation exposes two core challenges: 
C1. \emph{scalability across diverse proposal simulations}. Governance workflows vary substantially across DAOs in terms of proposal formats, execution pipelines, and contract architectures. Supporting large-scale mismatch analysis therefore requires a unified, project-agnostic simulation mechanism that can reproducibly derive execution traces across heterogeneous governance deployments.
C2. \emph{compatibility between descriptions and underlying execution traces}. Proposal descriptions are expressed in free-form natural language and capture high-level intent, while execution traces consist of structured, low-level function calls with diverse naming conventions and invocation patterns. Comparing these two different representations requires mapping between textual intent and heterogeneous execution semantics in a systematic manner.

However, even if these two challenges are addressed, verifiable proposal
mismatch is still not guaranteed.
Proposal descriptions are attacker-controlled inputs, and adversaries can
strategically craft language to manipulate automated reasoning.
As a result, a naive description-trace detector may be misled by vague or over-permissive phrasing.
Accordingly, we identify this common concern as: \emph{robustness against evasive or misleading descriptions}

An illustrative case is that the execution trace contains a privileged control-change action such as \texttt{transferOwnership()}.
In its original form, the proposal description focuses on benign marketing fund allocation and explicitly claims that no governance or control changes are involved, under which a verifier correctly marks the ownership transfer as not described. An attacker can evade this check by inserting a single, over-permissive sentence (e.g., vaguely stating that certain "management responsibilities may be adjusted
as needed"). Although this addition does not explicitly disclose any control transfer, its broad wording can be stretched to cover a wide range of privileged actions, leading a naive verifier to falsely accept the malicious execution as described.

\PP{Approach Overview}
To address the first challenge, we propose a \emph{lifecycle-based simulation} technique (\ding{182} in \autoref{fig:framework}). Our key observation is that governance proposal histories are publicly observable on-chain, and that proposals within the same DAO consistently reuse lifecycle-specific functions across the \texttt{propose}, \texttt{vote}, and \texttt{execute} stages. We therefore scan historical transactions to build a per-DAO \emph{governance profile} capturing eligible voters, timing parameters, and execution formats. For each new or live proposal, we combine its payload with the corresponding profile and drive it through the full lifecycle, a step we call \emph{live proposal simulation}. This yields reproducible, framework-agnostic execution traces across heterogeneous DAOs without privileged access (see \autoref{lifecycle-based-simulation}).


To address the second challenge, we propose a \emph{semantic-based DEMI guard} (\ding{183} in \autoref{fig:framework}) built on large language models (LLMs). Proposal description diversity renders rule-based keyword matching and traditional NLP methods impractical, and even a naive LLM query with a holistic YES/NO judgment misses a large fraction of DEMI attacks (46.6\% recall in our ablation, \autoref{sec:llm-ablation}). Our key design is therefore to follow an \emph{evidence-mapping} paradigm that mirrors human auditing: rather than requesting an overall verdict, we require the LLM to locate an explicit supporting span in the description for each executed action individually, grounding reasoning in concrete textual evidence rather than implicit coding knowledge (see \autoref{sec:consistendy_guard}).

To address robustness against evasive descriptions, we introduce a \emph{risk-aware robustness guard} (\ding{184} in \autoref{fig:framework}). Our key insight is that different execution actions require different levels of descriptive clarity: low-risk operations (e.g., metadata updates) tolerate coarse explanations, whereas high-risk actions (e.g., fund transfers or privilege changes) must be supported by explicit, purpose-aligned evidence. The guard enforces stricter evidence requirements for high-impact behaviors, rejecting vague, fabricated, or purpose-misaligned descriptions that would otherwise allow adversarial ambiguity to undermine mismatch verification (see \autoref{sec:robusness_guard}).

\subsection{Lifecycle-based Simulation}
\label{lifecycle-based-simulation}

\begin{figure*}[t]
    \centering
    \includegraphics[width=0.78\textwidth]{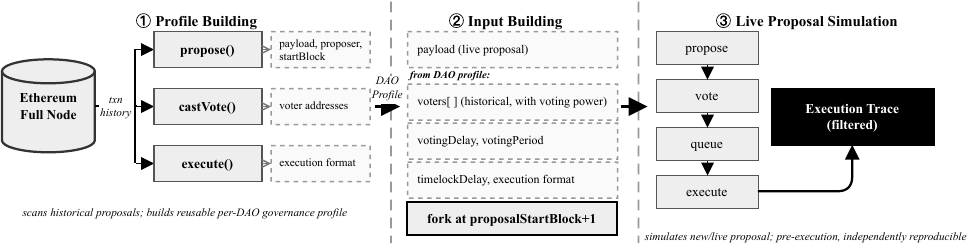}
    \caption{The three-stage lifecycle simulation pipeline that derives the execution trace.}
    \label{fig:data_aggregation}
\end{figure*}

To derive verifiable execution results without relying on privileged
administrator accounts, our design follows three phases
(\ding{182} in \autoref{fig:framework}, illustrated in \autoref{fig:data_aggregation}):
\ding{172}~\emph{profile building}, which scans historical on-chain data to
construct a reusable governance profile per DAO;
\ding{173}~\emph{input building}, which assembles simulation inputs for each
target proposal by combining its payload with the governance profile; and
\ding{174}~\emph{live proposal simulation}, which drives the proposal through
the full governance lifecycle and produces a filtered execution trace.

\noindent\textbf{\ding{172}~Profile Building.}
Driving a proposal through its full lifecycle requires governance parameters
and voter identities that are neither explicitly indexed nor standardized
across frameworks. Our key observation is that, despite interface-level
heterogeneity, the required information is consistently recorded on-chain
through proposal-related transactions. We therefore scan the transaction
history of each governor contract and, as illustrated in
\autoref{fig:data_aggregation}, categorize proposal-related calls into three
groups: \texttt{propose}, \texttt{castVote}, and \texttt{execute}.
From \texttt{propose} calls, we extract proposal payloads (targets, values,
calldatas), descriptions, proposers, governors, and start blocks.
From \texttt{castVote} calls, we recover voter identities and vote encodings.
From \texttt{execute} calls, we infer execution formats and callers.
To support diverse frameworks, we further normalize differences in proposal
identifier types (\texttt{uint} or \texttt{bytes32}) and function-signature
variants (\texttt{VoteHex}, \texttt{QueueHex}, and \texttt{ExecuteHex}).
Governance-level information (voter identities, execution formats, and
timing parameters) is aggregated per governor contract into a reusable
\emph{DAO Profile} indexed by \texttt{GovernorAddress}.
Proposal-specific data (payload, proposer, start block) is recorded separately
per proposal, ready to be combined with the DAO Profile during input building.

Voter identities illustrate why this reconstruction is necessary: invoking
\texttt{castVote} from arbitrary addresses is ineffective because most
addresses lack voting power. We therefore extract real voter addresses from
historical \texttt{castVote} callers and re-cast their votes during simulation,
reliably advancing proposals through the voting phase without any privileged
role.

\noindent\textbf{\ding{173}~Input Building.}
With the DAO Profile in hand, assembling simulation inputs for a specific
proposal is straightforward: the proposal's payload, proposer, and start block
are combined with the profile's voter addresses, timing parameters, and
execution format to form a self-contained input set.
A critical design choice is the simulation fork point: we fork the Ethereum
state at \texttt{proposalStartBlock + 1}, the block immediately following
proposal submission, before any votes have been cast or execution has occurred.
This ensures the simulation operates on a consistent pre-execution state
derived entirely from public on-chain data, and applies equally to historical
proposals and live pending proposals during their voting window.
Consequently, all required inputs are available at proposal submission
time, enabling our tool to issue warnings while the community can still
contest or reverse a malicious proposal.

\noindent\textbf{\ding{174}~Live Proposal Simulation.}
Directly simulating proposal execution is often infeasible in practice,
as many governance frameworks enforce privileged role checks or
state-dependent constraints that prevent arbitrary execution.
With the simulation inputs assembled, we instead drive each proposal through
the entire governance process, rather than invoking execution in isolation.

Our implementation, shown in \autoref{lst:lifecycle_simulation}, is built on
Foundry's \texttt{forge} testing framework.
\texttt{Forge} supports advancing block height and timestamp
(\texttt{roll} and \texttt{warp}) to satisfy governance timing constraints such
as voting delay, voting period, and timelock delay, and impersonating
externally owned accounts via \texttt{prank} to simulate proposer, voter, and
executor roles.
By combining these primitives, we drive a proposal through the canonical
\texttt{propose}--\texttt{vote}--\texttt{queue}--\texttt{execute} lifecycle
using only publicly observable on-chain data.


\begin{figure}[t]
    \centering
\begin{lstlisting}[style=simcode]
/* Inputs:
 *  RPC, forkBlock (= proposalStartBlock+1), governor G
 *  pid (if supported), proposer
 *  payload = (targets, values, calldatas, descriptionHash)
 *  voters[] (pre-collected, sufficient voting power)
 */
// fork at pre-execution state (proposalStartBlock+1)
fork(RPC, forkBlock);
(vd, vp)    = (G.votingDelay(), G.votingPeriod());
(TL, delay) = tryGetTimelockAndDelay(G); // optional; fallback if unavailable

// 1) advance to voting start, cast votes to pass the proposal
rollAndWrap(block.number + vd + 1, block.timestamp + vd + 1);
for v in voters {
    prank(v);
    castVoteVariants(G, pid, /*FOR=*/1);
}

// 2) advance to voting end, then queue
rollAndWrap(block.number + vp + 1, block.timestamp + vp + 1));
prank(proposer);
queueVariants(G, pid, payload, proposer);

// 3) advance to executable time, then execute
warp(block.timestamp + delay + 1);
executeVariants(G, pid, payload, proposer);
\end{lstlisting}
\caption{Live proposal simulation.}
\label{lst:lifecycle_simulation}
\end{figure}

The raw trace produced by the simulation is typically verbose.
We therefore apply a lightweight filtration step: (i) remove read-only calls
that do not modify on-chain state, (ii) discard subtraces triggered by
\texttt{delegatecall}, whose semantics largely duplicate the corresponding
external call paths, and (iii) exclude auxiliary segments introduced by
our simulation script that fall outside the proposal's actual
\texttt{execute()} context.
The remaining actions capture the proposal's essential state-changing and
value-moving behaviors, and serve as the execution-side input for downstream
mismatch analysis.

\subsection{DEMI Guard}\label{sec:consistendy_guard}

\begin{figure}[t]
    \centering
    \includegraphics[width=0.95\linewidth]{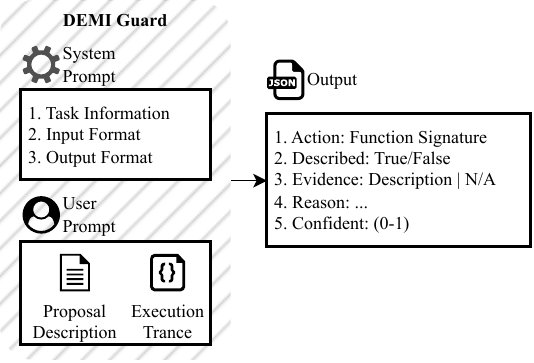}
    \caption{DEMI Guard.}
    \label{fig:consistency_guard}
\end{figure}

Semantic-based DEMI Guard (\ding{183} in \autoref{fig:framework}) is critical because governance decisions are made based on human-facing descriptions, while the actual effects are determined by on-chain execution. As discussed in \autoref{subsec:limitation} and \autoref{subsec:challenges-overview}, relying on manual inspection does not scale and offers weak protection for non-expert voters. At the same time, mismatch checking is inherently challenging because the forms differ between free-form natural language descriptions and structured execution behaviors, making direct comparison non-trivial.

Our key insight is that a mismatch does not require full semantic equivalence between descriptions and execution traces. Instead, it is sufficient to require action-level support: each state-changing or value-moving action in the execution trace should be explicitly supported by textual evidence in the proposal description. Based on this observation, we formulate mismatch detection as an \emph{evidence-mapping} problem. Given a set of extracted execution actions, we automatically identify minimal supporting spans in the description for each action. Actions without supporting evidence are flagged as potentially mismatched, enabling systematic and automated verification.

Concretely, the DEMI Guard takes as input a proposal description and the list of executed actions derived from the execution trace. For each action, it attempts to identify a corresponding sentence or phrase in the description that explicitly justifies the action. If no such evidence exists, the action is marked as \texttt{not\_described}. \autoref{fig:consistency_guard} shows the setting of prompts used for this evidence-mapping process. The output is a structured JSON object that records, for each action, the extracted evidence (if any), a binary decision, and a confidence score, along with an overall mismatch summary. This structured output serves as the basis for subsequent robustness checks and risk assessment.

\subsection{Robustness Guard}\label{sec:robusness_guard}

\begin{figure}[t]
\begin{lstlisting}[style=prompttpl]
[Knowledge (System Prompt)]
You are a DAO governance auditor with expertise in DeFi protocols, smart contracts, and on-chain governance semantics. You need to analyze a governance proposal description and find corresponding evidence for each executed DSL action. Remember the following:
  1. Mismatch definition: an action is "not described" if no sentence in the description explicitly justifies it.
  2. High-risk action taxonomy: <High-risk Taxonomy>
  3. For high-risk actions, evidence must cite an explicit address, amount, or named operation -- generic verbs alone (e.g., "update", "revoke") are insufficient.

[Detection Task (User Prompt)]
I am providing you with a governance proposal:
  1. Proposal description: <Description>
  2. Executed DSL actions from the on-chain trace: <Action List>

For each DSL action, find the sentence or phrase in the description that explicitly justifies it. Reason independently five times and report the most frequent result. Output as JSON:
{"action_mapping": [{"dsl_action": ...,
  "corresponding_text": "exact quote or null",
  "is_described": true/false, "is_high_risk": true/false,
  "confidence": 0.0-1.0, "reason": "brief explanation"}],
 "overall_consistency": {"described_actions": N,
  "total_actions": N, "consistency_rate": 0.0-1.0,
  "summary": "..."}}
\end{lstlisting}
\caption{Prompt template of DEMI Guard with Robustness design.}
\label{fig:robustness_guard}
\end{figure}

As discussed in \autoref{subsec:challenges-overview}, proposal descriptions are attacker-controlled inputs. Even when execution results are objectively derived (C1) and description-trace comparison is automated (C2), verifiable proposal mismatch is not guaranteed. Adversaries can strategically craft evasive or over-permissive language to manipulate automated reasoning, causing a naive mismatch verifier to falsely accept malicious executions as described. The evasive example in \autoref{subsec:challenges-overview} illustrates that DEMI alone is insufficient in adversarial settings.

Our key insight is that evasive descriptions succeed not by hiding malicious actions, but by exploiting uneven semantic clarity across different execution behaviors. Vanilla detection tends to apply uniform matching criteria, treating low-risk and high-risk actions equally. This allows attackers to use broad, weakly constraining language to justify critical operations under the same standard as benign ones. As a result, over-permissive descriptions can create sufficient semantic cover for dangerous behaviors without explicitly disclosing them. This observation suggests that robustness cannot be achieved by strengthening matching rules uniformly. Instead, effective defense requires calibrating the strictness of semantic justification according to the inherent risk of each executed action.

Robustness Guard (\ding{184} in \autoref{fig:framework}) operates as a second-stage validator on top of the DEMI Guard. It takes as input the proposal description and the action-evidence mappings produced by the DEMI Guard, and evaluates each mapping along two orthogonal dimensions: \emph{evidence clarity} (whether the evidence is explicit and action-specific) and \emph{action risk} (whether the action involves privileged control changes or value transfer). As illustrated in \autoref{fig:robustness_guard}, the Robustness Guard is implemented as a specificity constraint appended to the DEMI Guard prompt: for high-risk actions (token transfers, role/ownership changes, contract upgrades), the cited evidence must reference an explicit address, amount, or named operation, and generic verbs are rejected. Low-risk actions with clear evidence are accepted without this additional requirement, and intermediate cases are flagged with warnings. By enforcing action-calibrated evidence standards, the guard prevents adversarially ambiguous descriptions from satisfying $\mathrm{Cons}(\phi(d),\tau)$ and strengthens the robustness of verifiable proposal mismatch. Because this requirement targets a structural property of the task rather than any enumerated attack pattern, its robustness generalizes to unseen proposals in our red-team/blue-team study (\autoref{sec:eval-robustness}).

\section{DEMI Guard Configuration Analysis}\label{sec:llm-design}
\label{sec:llm-ablation}

The DEMI Guard involves two orthogonal design dimensions: \emph{prompt strategy} (what information structure is presented to the LLM and what judgment it is asked to make) and \emph{model selection} (which underlying model and temperature are used).
We validate each dimension independently on the 144 manually labeled proposals from the Tally dataset, keeping the other dimension fixed.
This two-stage design also keeps configuration selection separate from deployment-time performance reporting in \autoref{sec:eval-accuracy}. Concretely, the tables in this section evaluate the LLM detection component (evidence-mapping and its guard variants) on the 144-proposal subset, whereas \autoref{sec:eval-accuracy} reports the full deployed system, including the Robustness Guard and $N{=}5$ aggregation, on the 3,864-proposal labeled set; this is why the absolute numbers differ across the two.
Across both dimensions, evidence-mapping emerges as the optimal prompting paradigm, and its advantage over simpler approaches holds consistently across the models and vendors we test.

\subsection{Effect of Prompt Strategy}
\label{sec:llm-prompt}

\begin{table}[t]
\centering
\caption{Detection strategy comparison on 144 labeled proposals.}
\label{tab:llm-ablation}
\small
\setlength{\tabcolsep}{5pt}
\renewcommand{\arraystretch}{1.05}
\resizebox{\linewidth}{!}{
\begin{tabular}{llccc}
\toprule
\textbf{Method} & \textbf{Model / Variant}
  & \textbf{Prec.} & \textbf{Recall} & \textbf{F1} \\
\midrule
\multicolumn{5}{l}{\textit{Rule-based baselines (no LLM)}} \\
\midrule
All-function keyword match         & ---     & 60.6\% & 97.7\% & 0.748 \\
High-risk keyword match            & ---     & 63.6\% & 39.8\% & 0.490 \\
\midrule
\multicolumn{5}{l}{\textit{LLM-based strategies}} \\
\midrule
DeFiAligner-adapted~\cite{DeFiAligner_AFT24}  & GPT-4o  & 63.1\% & 93.2\% & 0.752 \\
Direct query (naive)               & GPT-4o  & 61.2\% & 46.6\% & 0.529 \\
Direct query (naive)               & GPT-4.1 & 74.1\% & 57.1\% & 0.645 \\
Evidence-mapping (EM)              & GPT-4o  & 83.9\% & 83.0\% & 0.834 \\
Evidence-mapping (EM)              & GPT-4.1 & 84.5\% & 85.7\% & 0.851 \\
\midrule
\textbf{DEMI Guard (ours)}         & \textbf{GPT-4o}  & \textbf{79.3\%} & \textbf{100\%} & \textbf{0.884} \\
\bottomrule
\end{tabular}
}
\end{table}

\autoref{tab:llm-ablation} compares six detection strategies across two tiers.
The two \emph{rule-based baselines} require no LLM: \emph{all-function keyword} flags a proposal when any function name from the execution trace is absent from the description text; \emph{high-risk keyword} restricts the same check to sensitive functions (transfers, ownership changes, upgrades, role grants).
The \emph{DeFiAligner-adapted} paradigm~\cite{DeFiAligner_AFT24} injects structured domain knowledge upfront: a taxonomy of inconsistency categories and their definitions, then asks the model to classify any inconsistencies it finds across the proposal as a whole, producing a typed, structured verdict.
\emph{Direct querying} (naive) dispenses with both taxonomy and structure: it simply asks the model whether the execution contains malicious or hidden behavior not mentioned in the description, leaving the model to decide, without guidance, what counts as evidence and how to weigh it.
\emph{Evidence-mapping} (EM) takes a fundamentally different approach: rather than asking for a global judgment, it requires the model to locate an explicit textual justification for each individual on-chain action before forming a verdict, shifting the burden from holistic impression to per-action accountability.
Finally, the \emph{DEMI Guard} augments EM with a specificity constraint: for high-risk actions (token transfers, role grants, ownership transfers), generic verbs such as ``revoke'' or ``update'' without an explicit address or amount are rejected as insufficient evidence.

Both rule-based baselines fail in opposite directions: all-function keyword over-flags at 60.6\% precision (97.7\% recall, F1=0.748), while high-risk keyword under-flags at 39.8\% recall (F1=0.490), confirming that surface-level lexical matching cannot capture description--execution semantics.
The DeFiAligner-adapted prompt achieves high recall (93.2\%) but collapses precision to 63.1\% (F1=0.752): without per-action anchoring, any perceived intent gap triggers a flag.
Direct querying drops recall to 46.6\%--57.1\%, missing roughly half of malicious proposals because the model accepts vague description language as sufficient justification when no per-action grounding is required.
Evidence-mapping raises precision to 83.9\%--84.5\% and recall to 83.0\%--85.7\% across both models (F1 gains of $+$30.5 and $+$20.6 over naive baselines).
The DEMI Guard closes the remaining gap by rejecting over-permissive evidence for high-risk actions, raising recall to 100\% at 79.3\% precision (F1=0.884).

\subsection{Effect of Model Selection}
\label{sec:llm-model}

\begin{table}[t]
\centering
\caption{Model and temperature sensitivity.}
\label{tab:llm-model}
\small
\setlength{\tabcolsep}{5pt}
\renewcommand{\arraystretch}{1.05}
\resizebox{\linewidth}{!}{
\begin{tabular}{llcccc}
\toprule
\textbf{Strategy} & \textbf{Model} & \textbf{Temp.}
  & \textbf{Prec.} & \textbf{Recall} & \textbf{F1} \\
\midrule
\multirow{2}{*}{Direct query (naive)} & GPT-4o-mini & 0.0 & 62.1\% & 46.6\% & 0.532 \\
                                      & GPT-4.1     & 0.0 & 74.1\% & 57.1\% & 0.645 \\
\midrule
\multirow{5}{*}{Evidence-mapping}     & GPT-4o-mini & 0.0 & 72.2\% & 88.6\% & 0.796 \\
                                      & GPT-4o-mini & 1.0 & 73.8\% & 86.4\% & 0.796 \\
                                      & GPT-4o      & 0.0 & 83.9\% & 83.0\% & 0.834 \\
                                      & GPT-4o      & 1.0 & 78.7\% & 79.5\% & 0.791 \\
                                      & GPT-4.1     & 1.0 & \textbf{84.5\%} & \textbf{85.7\%} & \textbf{0.851} \\
\bottomrule
\end{tabular}
}
\end{table}

\autoref{tab:llm-model} fixes the prompting strategy and varies model generation and temperature.
Within the evidence-mapping paradigm, F1 improves monotonically across generations: GPT-4o-mini (0.796) $\to$ GPT-4o (0.834) $\to$ GPT-4.1 (0.851).
The GPT-4o-mini to GPT-4o gain is driven primarily by precision (72--74\% $\to$ 79--84\%), reflecting a stronger ability to resolve subtle semantic correspondences between description text and on-chain function calls; GPT-4.1 further improves recall to 85.7\%, suggesting that the latest generation better handles undisclosed actions requiring multi-step contextual inference.
Temperature has a secondary effect within the same model tier: on GPT-4o, $t{=}0.0$ yields the highest precision (83.9\%) while $t{=}1.0$ is used in the final system to support the stochastic sampling that the Guard's multi-evidence aggregation benefits from; on GPT-4o-mini, both temperatures yield identical F1 (0.796), confirming robustness to this hyperparameter.

Beyond raw performance, the key observation is that the EM advantage over naive querying holds across the models we test rather than being an artifact of a single generation.
The F1 gap stands at $+$26.4 points on GPT-4o-mini (0.796 vs.\ 0.532) and $+$20.6 points on GPT-4.1 (0.851 vs.\ 0.645): while a more capable model does improve the naive baseline, it does not compensate for the absence of per-action grounding.
This indicates that the EM framework captures a structural requirement of the task that model scale alone does not satisfy. We test whether this advantage is specific to one vendor in \autoref{sec:llm-crossvendor}, and isolate the effect of self-consistency sampling in \autoref{sec:llm-selfconsistency}.

\subsection{Cross-Vendor Generalization}
\label{sec:llm-crossvendor}

\begin{table}[t]
\centering
\caption{Evidence-mapping vs.\ naive querying across eight models (three vendors), on the 144 labeled proposals.}
\label{tab:crossvendor}
\small
\setlength{\tabcolsep}{5pt}
\renewcommand{\arraystretch}{1.05}
\resizebox{\linewidth}{!}{
\begin{tabular}{lcccccc}
\toprule
\multirow{2}{*}{\textbf{Vendor / Model}} & \multicolumn{3}{c}{\textbf{Direct query (naive)}} & \multicolumn{3}{c}{\textbf{Evidence-mapping}} \\
\cmidrule(lr){2-4}\cmidrule(lr){5-7}
 & \textbf{Prec.} & \textbf{Recall} & \textbf{F1} & \textbf{Prec.} & \textbf{Recall} & \textbf{F1} \\
\midrule
OpenAI GPT-4o               & 61.2\% & 46.6\% & 0.529 & 83.9\% & 83.0\% & 0.834 \\
OpenAI GPT-4.1              & 74.1\% & 57.1\% & 0.645 & 84.5\% & 85.7\% & 0.851 \\
OpenAI GPT-5               & 71.5\% & 62.5\% & 0.667 & 60.8\% & 98.9\% & 0.753 \\
OpenAI GPT-5.6             & 70.4\% & 64.8\% & 0.675 & 76.4\% & 92.0\% & 0.835 \\
Anthropic Claude Sonnet 4.5 & 72.3\% & 78.8\% & 0.754 & 68.3\% & 98.5\% & 0.807 \\
Anthropic Claude Opus 4.8   & 72.9\% & 68.4\% & 0.706 & 78.6\% & 93.7\% & 0.855 \\
Anthropic Claude Fable 5    & 79.0\% & 35.1\% & 0.486 & 81.4\% & 85.7\% & 0.835 \\
Moonshot Kimi K3            & 75.9\% & 54.0\% & 0.631 & 62.0\% & 92.0\% & 0.741 \\
\bottomrule
\end{tabular}
}
\end{table}

To test whether the evidence-mapping advantage is specific to OpenAI models, we repeat the naive-vs-EM comparison across eight models spanning three vendors and several generations: OpenAI GPT-4o, GPT-4.1, GPT-5, and GPT-5.6; Anthropic Claude Sonnet~4.5, Claude Opus~4.8, and Claude Fable~5; and Moonshot Kimi~K3 (\autoref{tab:crossvendor}).\footnote{Dated snapshots are pinned where available (\texttt{gpt-4o-2024-08-06}, \texttt{gpt-4.1-2025-04-14}); the other identifiers (\texttt{gpt-5}, \texttt{gpt-5.6-luna}, \texttt{claude-sonnet-4-5}, \texttt{claude-opus-4-8}, \texttt{claude-fable-5}, \texttt{kimi-k3}) are the providers' 2026 identifiers, for which no dated snapshot is published.}
The pattern is consistent across every model tested: evidence-mapping sustains recall between 83.0\% and 98.9\%, whereas naive direct querying on the same models is strongly model-dependent, ranging from 35.1\% (Claude Fable~5) to 78.8\% (Claude Sonnet~4.5). On every model, restructuring the same query as per-action evidence mapping raises recall, by as much as $+$50 points (Claude Fable~5: 35.1\% $\to$ 85.7\%).
The absolute F1 of EM varies across models (0.74--0.86), driven by precision (60.8--84.5\%) rather than recall (83.0--98.9\%): several models flag a few additional borderline proposals, trading precision for the near-perfect recall that a recall-first detector prioritizes.
The property that matters for our setting, namely not missing malicious proposals, is thus preserved across the vendors and generations we test, indicating that the benefit of task decomposition and per-action grounding holds broadly rather than being specific to a single provider.

\subsection{Effect of Self-Consistency}
\label{sec:llm-selfconsistency}

\begin{table}[t]
\centering
\caption{Self-consistency ablation ($N$ passes, majority vote) on the 144 labeled proposals.}
\label{tab:selfconsistency}
\small
\setlength{\tabcolsep}{6pt}
\renewcommand{\arraystretch}{1.05}
\begin{tabular}{ccccc}
\toprule
\textbf{$N$} & \textbf{F1} & \textbf{Prec.} & \textbf{Recall} & \textbf{Cost / prop.} \\
\midrule
1 & 0.812 & 71.9\% & 93.2\% & \$0.016 \\
3 & 0.843 & 74.1\% & 97.7\% & \$0.048 \\
5 & 0.839 & 73.5\% & 97.7\% & \$0.080 \\
7 & 0.835 & 72.9\% & 97.7\% & \$0.112 \\
\bottomrule
\end{tabular}
\end{table}

The Guard aggregates $N$ independent reasoning passes by majority vote (\autoref{fig:robustness_guard}). We sweep $N\in\{1,3,5,7\}$ on the 144 labeled proposals (\autoref{tab:selfconsistency}) to isolate the contribution of this sampling from that of the evidence-mapping structure itself.
Evidence-mapping already achieves F1${=}0.812$ at $N{=}1$ (precision 71.9\%, recall 93.2\%), far above the naive baseline (0.529 on the same set), so its advantage is not an artifact of majority voting.
Self-consistency then adds a small, saturating gain: recall rises from 93.2\% ($N{=}1$) to a stable 97.7\% for every $N{\ge}3$, and F1 plateaus near 0.84 (0.843/0.839/0.835 at $N{=}3/5/7$) with no measurable benefit beyond $N{=}3$.
The 97.7\% figure is not a small-sample coincidence: the same two borderline malicious proposals missed at $N{=}1$ are recovered in \emph{every} multi-pass run ($N{\ge}3$), so the improvement reflects variance reduction on genuinely ambiguous cases rather than a lucky draw.
Because the passes are independent they execute in parallel, so end-to-end latency stays close to a single pass ($\sim$8.8s) at every $N$, whereas a serial implementation would scale as $N{\times}$8.8s; the dollar cost scales linearly ($\$0.016$ per pass at GPT-4o rates: $\$0.048$ at $N{=}3$, $\$0.080$ at $N{=}5$, $\$0.112$ at $N{=}7$).
The deployed Guard uses $N{=}5$ for a small stability margin; $N{=}3$ attains essentially the same accuracy at lower cost.

\section{Evaluation}\label{sec:evaluation}

We conduct an empirical evaluation to validate whether our design addresses the key challenges underlying verifiable description--execution mismatch (DEMI). The experiments are organized around the three challenges introduced earlier in \autoref{subsec:challenges-overview}, with each subsection evaluating one corresponding capability of our system. 

First, to assess whether we can enable execution-result derivation across heterogeneous governance frameworks (C1), we measure the \emph{coverage} of our simulation pipeline: how many DAO projects can be supported end-to-end and, within those projects, how many distinct proposals can be independently simulated to obtain execution results. 
Second, to evaluate whether our approach makes pure-text descriptions comparable to structured execution behaviors (C2), we measure the \emph{accuracy} of our DEMI detector on real-world proposals, reporting standard confusion-matrix-based classification metrics. 
Third, because proposal descriptions are attacker-controlled and may be crafted to evade automated checks (C3), we evaluate the \emph{robustness} of our system. Beyond an ablation of the \emph{robustness guard} on historical evasions, we stress-test it against an adaptive adversary through an iterative red-team/blue-team study varying the attacker's knowledge and capability, a held-out generalization test on disjoint proposals, and a prompt-injection stress test with input-isolation defenses. Together these experiments provide an end-to-end validation of our design.

\subsection{Simulation Coverage}
\label{sec:eval-coverage}

\begin{figure}[t]
    \centering
    \includegraphics[width=\linewidth]{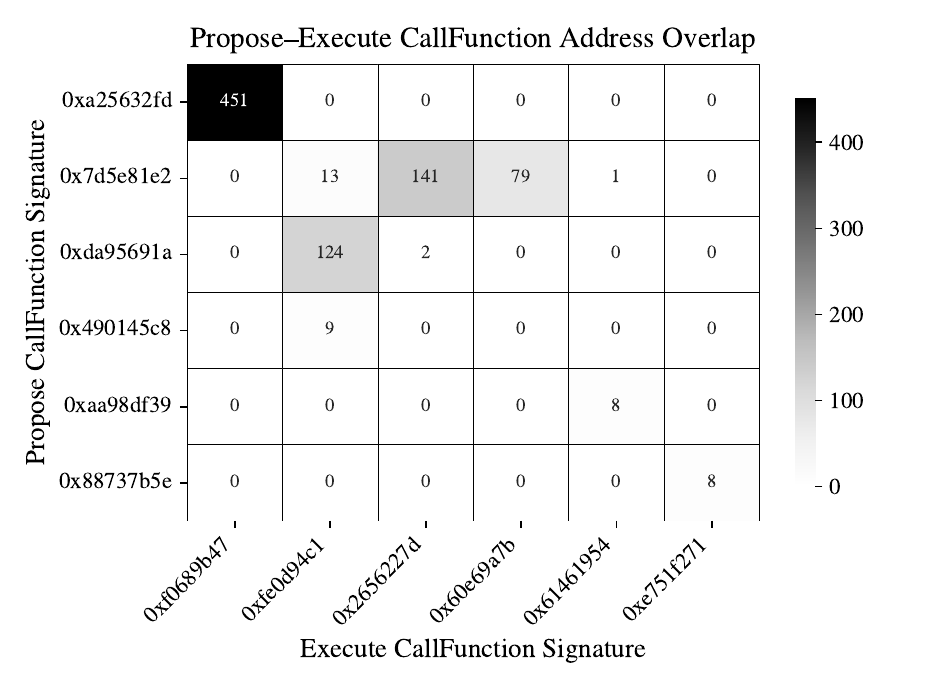}
    \caption{Function-signature usage across the proposal lifecycle.}
    \label{fig:APIs_class}
\end{figure}

\begin{table}[t]
\centering
\small
\setlength{\tabcolsep}{6pt}
\renewcommand{\arraystretch}{1.05}
\resizebox{\linewidth}{!}{
\begin{tabular}{lcccc}
\toprule
\textbf{Platform} &
\textbf{Description} &
\textbf{Payload} &
\textbf{Exec. Result} &
\textbf{Authentic} \\
\midrule

Snapshot
& \checkmark &   &   &   \\

Tally
& \checkmark & \checkmark & \checkmark$^{\dagger}$ &   \\

Project-native portals
& \checkmark & \checkmark &   &   \\

\midrule
\textbf{Ours}
& \checkmark & \checkmark & \checkmark & \checkmark \\

\bottomrule
\end{tabular}
}
\caption{Proposal execution visibility and authenticity across governance platforms.
$\dagger$ Tally results are optional and proposer-supplied, without independent reproducibility guarantees.}
\label{tab:coverage-comparison}
\end{table}

\begin{table}[t]
\centering
\small
\setlength{\tabcolsep}{6pt}
\begin{tabular}{lrr}
\toprule
\textbf{Metric} & \textbf{Supported} & \textbf{Coverage} \\
\midrule
DAO-level coverage      & 332 / 358 & 92.7\% \\
Proposal-level coverage & 9,098 / 10,190 & 89.3\% \\
\bottomrule
\end{tabular}
\caption{Coverage of our simulation pipeline over active DAOs.}
\label{tab:sim-coverage}
\end{table}

\PP{Setup}
We construct our evaluation dataset via a whole-chain scan that identifies 3,681 governance-related contracts on Ethereum, from which we select \emph{active}, \emph{community-auditable}, and \emph{vote-enabled} DAO deployments using three criteria:
\textit{(1) Active governance usage}, retaining only deployments with a history of executed proposals (filtering out contracts that expose governance interfaces but never execute community-approved actions);
\textit{(2) Community-auditable governance logic}, retaining only contracts verified on Etherscan (i.e., open-source deployments), since our goal of verifiable mismatch relies on public inspection;
\textit{(3) Vote-enabled governance frameworks}, excluding an early administrator-oriented template identified by the \texttt{propose}/\texttt{execute} signature pair \texttt{0xa25632fd}/\texttt{0xf0689b47} via function-signature analysis over the proposal lifecycle (\autoref{fig:APIs_class}).
After applying these criteria, we obtain 358 DAO deployments covering 10,190 executed proposals. As a point of comparison, Tally~\cite{Tally_explore} indexes 154 DAOs at the time of our measurement, highlighting the broader coverage enabled by a platform-independent, whole-chain view.

\PP{Results}
\autoref{tab:coverage-comparison} shows that existing governance platforms
primarily expose proposal artifacts without providing independently verifiable
execution results; in contrast, our system is the only one that derives
authentic execution traces. As reported in \autoref{tab:sim-coverage}, our
simulator supports 332 out of 358 active DAOs (92.7\%) at the DAO level and
successfully derives execution results for 9,098 out of 10,190 executed
proposals (89.3\%) at the proposal level.

\PP{Coverage limitations.}
The 26 uncovered DAOs and 1,092 uncovered proposals stem from on-chain
reverts, external oracle dependencies unavailable at the fork block, and
non-standard voting logic; none indicates silent mis-simulation.
Our current scope is Ethereum L1 Governor-pattern DAOs. Cross-chain governance
(L2-native timelocks, bridge-based voting) remains future work.

\subsection{Detection Accuracy}
\label{sec:eval-accuracy}

To evaluate whether our approach makes pure-text proposal descriptions
systematically comparable to structured execution behaviors (C2),
we measure the detection accuracy on a large, real-world governance dataset
indexed by Tally~\cite{Tally_explore}.
Tally aggregates governance metadata and proposal records across
154 DAOs deployed on Ethereum.

\PP{Setup}
We evaluate on \emph{historically executed} proposals to obtain ground-truth mismatch labels; this does not limit applicability, since our simulation forks the chain at \texttt{proposalStartBlock + 1} (a pre-execution state) and yields identical results for pending proposals.
In total, we collect 3,864 proposals, of which
893 are labeled \emph{mismatched} and 2,971 \emph{matched}.
Labels were produced by two domain experts following a dual-annotator protocol
(inter-annotator agreement 98.5\%, Cohen's $\kappa=0.958$)~\cite{landis1977measurement}; the 58 discrepant
cases were resolved through consensus review of on-chain transaction history,
community records, and financial loss evidence
\ifarxiv (see \autoref{sec:appendix_ground_truth} for the full methodology).\fi
We evaluate our detector on the full dataset using
stratified 4-fold cross-validation to assess robustness and
generalization under realistic, skewed governance distributions.

\begin{table}[t]
\centering
\small
\setlength{\tabcolsep}{6pt}
\renewcommand{\arraystretch}{1.05}
\resizebox{\linewidth}{!}{
\begin{tabular}{lccc}
\toprule
\textbf{Configuration} & \textbf{Precision} & \textbf{Recall} & \textbf{F1} \\
\midrule
w/o Robustness Guard & 79.8\% $\pm$1.5\% & 82.4\% $\pm$2.1\% & 0.810 $\pm$0.018 \\
\textbf{Full system (ours)} & \textbf{81.7\% $\pm$1.1\%} & \textbf{98.3\% $\pm$0.6\%} & \textbf{0.892 $\pm$0.008} \\
\bottomrule
\end{tabular}
}
\caption{Detection accuracy under stratified 4-fold cross-validation.}
\label{tab:detection-accuracy}
\end{table}

\PP{Results.}
\autoref{tab:detection-accuracy} reports detection accuracy under 4-fold
cross-validation for two configurations: with and without the Robustness Guard.
Without the guard, the system achieves 79.8\% precision and 82.4\% recall.
Enabling the Robustness Guard raises recall to 98.3\% with a negligible
change in precision (81.7\%), yielding an 8-point F1 improvement (0.810 to 0.892).
The low variance across folds confirms that these results are stable across
different data partitions.
High recall is a critical property in governance settings where accepting a
malicious proposal causes irreversible damage.

\PP{Operating point and deployment role.}
The 81.7\% precision reflects a deliberate recall-first operating point, not a
ceiling imposed by the detector. Sweeping the decision threshold on the labeled
set traces a precision--recall frontier on which precision can be traded up only
by sacrificing recall; at our chosen point recall is near-saturated (98.3\%),
which is the priority when a single missed proposal can drain a treasury
irreversibly while a false alarm merely triggers a human review. We therefore
position the system as a \emph{triage} layer for governance platforms and
security firms: it surfaces the small set of proposals that warrant manual
scrutiny before execution, rather than an autonomous accept/reject oracle. Under
this role the residual false positives (roughly one in five flagged proposals)
are inexpensive, since each is adjudicated by a reviewer in the pre-execution
window, whereas a false negative is not recoverable.
Inspecting these false positives, they concentrate on the under-specified
``irresponsible'' proposals characterized in \autoref{sec:hygiene}: benign
descriptions that omit execution-relevant detail, which the guard flags as
unverifiable even though no adversarial intent is present. Surfacing such proposals
for human review is the intended behavior rather than a true error.

\subsection{Robustness Evaluation}
\label{sec:eval-robustness}

To evaluate whether our system is robust to evasive or misleading proposal descriptions (C3), we compare detection performance with and without the Robustness Guard, using the full-corpus results from \autoref{sec:eval-accuracy} (\autoref{tab:detection-accuracy}) for aggregate metrics and the Tally subset ablation from \autoref{sec:llm-ablation} (\autoref{tab:llm-ablation}) for per-configuration detail.

\PP{Ablation results.}
As shown in \autoref{tab:detection-accuracy}, removing the Robustness Guard
drops recall from 98.3\% to 82.4\% while precision changes minimally,
confirming that the guard eliminates false negatives on evasive proposals
rather than filtering false alarms.
The per-configuration breakdown in \autoref{tab:llm-ablation} shows this
recall gap is consistent across model and temperature choices, establishing
that the guard's specificity constraint is essential for adversarial
deployment rather than an artifact of any single configuration.

\ifarxiv
\PP{Adversarial case analysis.}
Among the 69 confirmed malicious proposals, we identify three recurring
evasion patterns (detailed in \autoref{sec:appendix_adversarial}):
action substitution, legitimacy laundering, and partial disclosure, each of
which the Robustness Guard resolves by rejecting over-permissive evidence that
base EM accepts.
\fi

\begin{figure}[t]
    \centering
    \includegraphics[width=\linewidth]{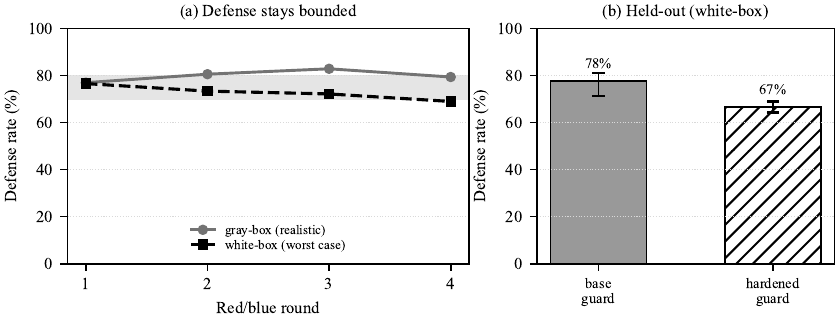}
    \caption{Adversarial robustness (3 trials; mean, whiskers${=}$range). (a) Defense stays bounded near $70$--$80\%$ across rounds for both a gray-box and a worst-case white-box attacker. (b) On held-out proposals, the base guard (no learned rules) matches or exceeds the hardened guard, so the robustness appears structural.}
    \label{fig:redblue}
\end{figure}

\PP{Adaptive red-team/blue-team evaluation.}
To assess robustness against an adversary aware of our detector, we conduct an
\emph{iterative} red-team/blue-team study over three independent trials, varying
both what the attacker knows and how capable it is. In each round a red-team LLM
rewrites the 84 detected-malicious descriptions to evade the \emph{current} guard
while preserving the malicious execution; a blue-team LLM then distills the round's
\emph{stealthy} evasions (those that evade the guard yet would still mislead voters)
into new rules. Each trial runs four rounds, and the red-team attacker, the guard,
and the voter judgment are all instantiated with GPT-4o, so the adversary has
capability equal to the detector. Whether an evasion would still mislead voters is
decided by a separate temperature-0 voter model that inspects the description
against the true on-chain actions; an attack counts as defended once it is either
flagged by the guard or surfaces the high-risk recipient or amount to that voter.
We study two knowledge settings (\autoref{fig:redblue}): a \emph{white-box}
attacker handed our exact guard prompt and high-risk taxonomy (a pessimistic worst
case), and a \emph{gray-box} attacker that knows only the published
method: evidence-mapping with extra scrutiny on high-risk actions, but not the
exact prompt, category list, or thresholds (the realistic case for a reader of this
paper).

No content-based detector rules out adaptive evasion entirely, so the useful
questions are how much an adversary can achieve and how much an automated
defender can add. Two findings answer them. \emph{First, the red team's impact is bounded.} Across
both settings, defense stays near $70$--$80\%$ in every round (\autoref{fig:redblue}a),
so even our strongest white-box attacker, holding our exact detector, lands a
successful attack in only about $30\%$ of attempts (about $20\%$ for the realistic
gray-box attacker). \emph{Second, an automated blue team did not meaningfully improve
on this.} Its validation-gated hardening lifts defense only marginally on the very
attacks it is tuned against, and that gain does not transfer. In a held-out test (we
split the 84 proposals in half by index, harden the guard on one half, and attack the
other half over three trials), the base guard with \emph{no} learned rules defends
$77.8\%$ of held-out white-box attacks, matching or exceeding the hardened guard
($66.7\%$), whose held-out defense stays within $3.2$ points of its defense on the
tuning set (\autoref{fig:redblue}b).

This is a statement about the design rather than a limitation of the hardening:
because a white-box adversary can route around any specific rule, the robustness
that remains resides in the evidence-mapping structure and the guard's high-risk
disclosure requirement, a heuristic that generalizes to proposals it was never
tuned on. We report this as an empirical observation under our procedure, not as
proof that no automated defender could help. These figures also assume the
attacker holds our detector; a realistic proposer faces the full system at
$98.3\%$ recall (\autoref{sec:eval-accuracy}), so we regard the guard as robust
for pre-execution triage.

\PP{Prompt-injection robustness.}
Because proposal descriptions are attacker-controlled text fed to an LLM, we
stress-test prompt injection: we embed four adversarial instruction templates
(direct override, forged security-council approval, fabricated blanket
authorization, and lenient-reviewer role-play) into the malicious descriptions.
Against an undefended guard prompt, injection flips the verdict to
``consistent'' for up to 60\% of cases with a single template and 68\% for at
least one template. Prompt injection is a general vulnerability of any LLM that
consumes untrusted input~\cite{greshake2023injection} rather than a weakness
specific to our design, so we apply standard input-isolation
defenses~\cite{hines2024spotlighting}: the description is placed in a
delimited untrusted block and the model is instructed to treat its contents
strictly as data, never as instructions. Under these defenses, injection success
falls to 4\% (a single residual case). We adopt these defenses in the deployed
guard.

\section{Practical DEMI Vulnerabilities}\label{sec:practical_attack}

To show practical impact beyond controlled benchmarks, we analyze real-world governance at scale. Across 10,190 executed proposals, our pipeline flags 671 as mismatched; manual review confirms 544, while the remaining 127 are ambiguous cases whose intent would require project-side confirmation. This deployment-scale scan is distinct from the 3,864-proposal labeled set used in \autoref{sec:eval-accuracy}. The 544 confirmed cases fall into two categories with markedly different economic exposure.

In total, we identify 69 proposals that exhibit clear malicious intent.
Within this set, 9 proposals involve privileged control changes and large-value fund transfers that were
capable of causing substantial and irreversible losses~\cite{Blocked_Loot_Attack_2026, WEN_attack, Tally_PEPE_CASH, Risy_security, Arbix_attack, Tally_Fei, Reflexer_Ungovernor_attack, Tally_Aggregated, Signata_attack}, as summarized in \autoref{tab:proposal_inconsistency_cases}.
The remaining 60 malicious proposals follow the same adversarial
patterns but are associated with smaller treasuries or limited execution scope,
leading to lower realized or attempted losses. 

Second, most confirmed cases (475 proposals) are irresponsible governance actions, where descriptions are materially mismatched with execution but no explicit exploit follows. Though rarely causing immediate loss, they reflect a pervasive governance-quality problem that misleads voters, obscures execution intent, and erodes trust.

Together these findings establish proposal mismatch as a systemic, security-critical issue: a few cases cause severe financial losses, while a much larger population of irresponsible proposals erodes everyday governance transparency, motivating automated verification rather than ad-hoc vigilance.

\begin{table}[t]
\centering
\small
\setlength{\tabcolsep}{5pt}
\renewcommand{\arraystretch}{1.05}
\resizebox{\linewidth}{!}{
\begin{tabular}{lcc}
\toprule
\textbf{DAO} & \textbf{Key Execution} & \textbf{Category} \\
\midrule
Fei DAO~\cite{Tally_Fei} & \texttt{changeProxyAdmin()} & Control Escalation \\
Pepe Cash~\cite{Tally_PEPE_CASH_779} & \texttt{transfer(12.62T PCASH)} & Fund Drain \\
Aggregated Finance~\cite{Tally_Aggregated} & \texttt{transfer(90 ETH)} & Fund Drain \\
Reflexer Ungovernor~\cite{Reflexer_Ungovernor_attack} & \texttt{transferERC20()} & Fund Drain \\
Signata DAO~\cite{Signata_attack} & \texttt{approve()}, \texttt{transfer()} & Fund Drain\\
Loot DAO~\cite{Blocked_Loot_Attack_2026} & \texttt{transfer(477 ETH)} & Fund Drain \\
Risy DAO~\cite{Risy_security} (and others) & Privileged state changes & Hygiene Issues \\
\bottomrule
\end{tabular}
}
\caption{Real-world DEMI attack cases confirmed.}
\label{tab:proposal_inconsistency_cases}
\end{table}

\subsection{Zero-day DEMI Attacks}

From the 69 malicious proposals in our dataset, we focus on the 9 previously unreported cases that we disclosed to and had confirmed by the project parties (\autoref{tab:proposal_inconsistency_cases}). They span distinct attack goals, from covert governance control escalation to direct value extraction, while sharing one weakness: a divergence between human-facing descriptions and actual execution behavior.

\PP{Control Escalation}
Here a financially plausible description conceals a privileged governance operation. The following Fei DAO case shows how a standard DeFi transaction narrative can mask a proxy admin reassignment granting the attacker full protocol control.

\begin{mdframed}[nobreak=true, frametitle={Case \#1 --- Fei DAO, Proposal-404~\cite{Tally_Fei}}, frametitlebackgroundcolor=black!75, frametitlefont=\small\bfseries\color{white}, frametitleaboveskip=3pt, frametitlebelowskip=3pt, linecolor=black, linewidth=0.5pt, innerleftmargin=6pt, innerrightmargin=6pt, innertopmargin=4pt, innerbottommargin=4pt]
\small
\textbf{Description:}
\begin{lstlisting}[style=quote]
"The proposal aims to enable the protocol to transfer its 34,038 vlAURA tokens to Fishy in exchange for 94,000 DAI through an over-the-counter (OTC) trade arrangement."
\end{lstlisting}
\textbf{Our simulation:} \;\texttt{changeProxyAdmin(newAdmin)}\\
\textbf{Mismatch:} Description claims a routine asset swap; execution reassigns proxy admin rights, transferring governance control to an attacker-controlled address.\\
\textbf{Impact:} Attempted governance takeover of a live DAO.
\end{mdframed}

The deception is structural: the proposal mirrors a previously-approved Fei DAO trade with the same counterparty, presenting itself as a routine OTC operation voters had already endorsed. Beneath that narrative, the calldata invokes \texttt{changeProxyAdmin} rather than a token transfer, smuggling a governance takeover into an apparent repeat vote.

\PP{Fund Drain}
The most prevalent category involves direct fund extraction, in two variants: social camouflage mimicking legitimate consensus, and outright omission relying on governance fatigue.

The following Pepe Cash DAO case illustrates camouflage: it adopts every convention of a legitimate action, citing prior discussion, named wallets, and precise amounts, yet the actual recipient in the trace is fabricated.

\begin{mdframed}[frametitle={Case \#2 --- Pepe Cash DAO, Proposal-779~\cite{Tally_PEPE_CASH_779}}, frametitlebackgroundcolor=black!75, frametitlefont=\small\bfseries\color{white}, frametitleaboveskip=3pt, frametitlebelowskip=3pt, linecolor=black, linewidth=0.5pt, innerleftmargin=6pt, innerrightmargin=6pt, innertopmargin=4pt, innerbottommargin=4pt]
\small
\textbf{Description:}
\ifarxiv
\begin{lstlisting}[style=quote]
"After initial preliminary voting within the community discussions on Discord, it was proposed to redistribute the $PCASH tokens in 'Claim Safe' address 0x62328E88f31aF0b6EF8b77Fb6ba069dD46B2C128. The consensus among voters was that this proposal should be approved on the chain to advise developers to complete the process formally. It was also agreed that 66% or 2/3 of the wallet holdings should be sent to the marketing wallet 0x9c446bb0195d0f2878960D4e38c986Bd693bFa7E. The total amount to be transferred is 4,026,000,000,000 (four trillion, twenty-six billion) to the marketing wallet."
\end{lstlisting}
\fi
\textbf{Our simulation:} \;\texttt{transfer(0xfd63...,\ 12.62T\ PCASH)}\\
\textbf{Mismatch:} Description names a specific marketing wallet as recipient; execution transfers to an attacker-controlled address \texttt{0xfd63...} with no relation to the stated destination.\\
\textbf{Impact:} Full token extraction; proceeds laundered via Tornado Cash.
\end{mdframed}

The Aggregated Finance cases represent omission: no meaningful description at all, relying purely on voter inattention to pass a large-value transfer.

\begin{mdframed}[nobreak=true, frametitle={Case \#3 --- Aggregated Finance, Proposal-7642~\cite{Tally_Aggregated}}, frametitlebackgroundcolor=black!75, frametitlefont=\small\bfseries\color{white}, frametitleaboveskip=3pt, frametitlebelowskip=3pt, linecolor=black, linewidth=0.5pt, innerleftmargin=6pt, innerrightmargin=6pt, innertopmargin=4pt, innerbottommargin=4pt]
\small
\textbf{Description:} \texttt{"a"} \textit{(verbatim; single character)}\\[3pt]
\textbf{Our simulations:} \;\texttt{transfer(90 ETH)}; \;\texttt{transfer(477 ETH)}\\
\textbf{Mismatch:} Description provides no justification; large-value transfers executed directly.\\
\textbf{Impact:} Fund extraction; passed vote despite absent narrative.
\end{mdframed}

These cases show that proposal mismatch is a systemic governance risk rather than isolated misuse, spanning subtle narrative-driven attacks and overt execution abuses. Being previously unreported yet project-confirmed, they demonstrate that existing platforms and manual review cannot surface such risks before execution, which scalable pre-execution mismatch analysis addresses.

\subsection{Governance Hygiene Issues}\label{sec:hygiene}

Beyond malicious proposals, a large body of \emph{irresponsible} ones share a root cause: \emph{under-specified descriptions}. Mismatch here stems not from adversarial intent but from poor hygiene: administrators treat proposals as lightweight operational tools (e.g., "governance-as-git"), omit execution-relevant details, or rely on implicit context the community does not share. Even when benign, such proposals widen the verification gap and condition communities to accept ambiguous narratives.

\PP{Under-specified descriptions in practice.}
We observe two recurring manifestations. First, some proposals use extremely short or test-like text (effectively empty) while the payload executes multiple on-chain actions; this pattern appears in test/low-stakes DAOs~\cite{Tally_3T_Governance,Tally_Flux,Tally_FRICTIONLESS,Tally_Joseon} and in larger projects where governance is treated as an administrative change-log~\cite{Tally_Idle,Tally_Ondo,Reflexer_Ungovernor_attack,Tally_sepolia_security_council}. Second, ``update''-style proposals whose execution includes privileged state transitions or value movements (e.g., transfers, administrative role changes) without disclosing the operational semantics (e.g., Radworks~\cite{Tally_Radworks} proposals invoking \texttt{setPendingAdmin}). These may be operationally legitimate yet remain mismatched from a voter's perspective, since the description does not constrain what execution will do.

Such non-malicious mismatches reveal a governance-quality problem: ambiguity normalizes trust-based voting and lets malicious proposals hide among noisy submissions, reinforcing the need for verifiable execution evidence and automated mismatch checking.

\section{Mitigation Discussion}\label{sec:discussion}


DEMI is a systemic risk rooted in the separation between human-facing descriptions and low-level execution semantics. We recommend safeguards at the protocol and workflow levels:
(i) \emph{execution previews}, mandating reproducible execution traces or effect summaries before voting, reducing reliance on proposer-supplied artifacts;
(ii) \emph{structured review workflows}, requiring explicit review stages, minimum discussion periods, or audit-conditioned quorum thresholds for high-impact proposals; and
(iii) \emph{risk-signal surfacing}, automatically flagging vague justifications for high-risk actions, inconsistent evidence, or excessive execution privileges.
Our system serves as a lightweight auditing layer supporting (i) and (iii), and reduces the cost of (ii) by automating mismatch checking.

\PP{Real-world acknowledgement.}
Among the 69 malicious proposals we surfaced, 9 are previously unreported vulnerabilities that we disclosed to and had confirmed by the corresponding project parties. The RisyDAO administrators confirmed our report, issued a security bounty, and promptly deployed governance upgrades~\cite{Risy_security, PolygonScan_bounty}. For inactive or legacy projects, cases were verified through security firms and direct communication with project teams via public channels such as Discord. These responses reflect recognition of DEMI as a practical threat.

\section{Related Work}
Prior work studies DAO security from empirical and system angles. Several efforts analyze real-world governance attacks and human factors~\cite{DAO_security_DeFiWorkshop,Good_SoK_for_DAO}, proposal quality at scale~\cite{DAO_process_explain_arxiv,DAO_Governance_security_review_tosem}, decentralization via Voting-Bloc Entropy~\cite{VBE_Usenix26}, and cryptographic verification frameworks~\cite{DAV_INES24}; complementary work addresses Web3 fraud~\cite{NFT_Rug_Pull} and governance mechanism design~\cite{DAO_Whale_colusion,Diamond_template_for_DAO}. None offers automated, execution-grounded mismatch verification. The closest work, DeFiAligner~\cite{DeFiAligner_AFT24}, uses symbolic analysis and LLMs to flag inconsistencies between documentation and deployed code; in contrast, governance proposals require full lifecycle simulation for a verifiable trace, and the holistic prompt of~\cite{DeFiAligner_AFT24} yields only 63.1\% precision in our task (\autoref{sec:llm-prompt}).

\section{Conclusion}
We present the first framework for verifiable description--execution mismatch (DEMI) detection in DAO governance, where a human-facing description that voters rely on can diverge from the on-chain payload that alone determines effects. Without privileged access, our framework reconstructs an authentic execution trace for each proposal, checks it against the description action by action, and demands explicit justification for high-risk operations. It simulates 89.3\% of proposals, detects mismatches at 81.7\% precision and 98.3\% recall, generalizes across eight LLMs from three vendors, withstands the majority of adaptive attacks even from an adversary that knows the detector, and surfaces 9~previously unreported, project-confirmed vulnerabilities across 10,190 proposals. We hope it serves as a pre-execution security primitive, so that governance decisions rest on what a proposal will actually do rather than only on what it claims.

\begin{acks}
Kangjie Lu, Bowen Cai and Weiheng Bai were supported in part by NSF awards
CNS-2045478, CNS-2106771, and CNS-2247434. Any opinions, findings, conclusions
or recommendations expressed in this material are those of the authors and do
not necessarily reflect the views of NSF.
\end{acks}

\bibliographystyle{sty/ACM-Reference-Format}
\bibliography{p}

\appendix
\normalsize
\section{Ethical Considerations}

\ifarxiv
Our study analyzes DAO governance proposals and execution traces to understand
and mitigate risks arising from proposal mismatch. All proposals discussed in
this paper correspond to historical on-chain events that have already occurred
or to proposals that were ultimately blocked or rendered ineffective. We do not
disclose any currently executable payloads, private keys, privileged accounts,
or actionable exploits that could be directly reused to compromise live
governance systems. Moreover, our system does not automate proposal creation or
attack execution; it is designed solely for retrospective analysis and
defensive detection.

We follow a responsible disclosure process for cases involving real-world
impact. Prior to publication, we made reasonable efforts to contact affected
project teams or governance maintainers when identifiable. In several
instances, including Risy DAO~\cite{Risy_security}, project teams acknowledged
the issues, deployed governance updates, and, in some cases, provided bug
bounties in recognition of the report~\cite{PolygonScan_bounty}. All disclosed
incidents either occurred in the past or no longer pose an immediate risk. To
the best of our knowledge, the disclosures in this paper do not introduce new
economic harm to the affected projects or their communities.
\fi

\section{Compliance with the Open Science Policy}

\ifarxiv
To support transparency, reproducibility, and independent verification, we
release a complete artifact bundle covering all components needed to reproduce
the empirical results in this paper.

\PP{Artifact contents.}
The release includes: (i) the simulation pipeline (lifecycle replay, governance
profile builder, on-chain trace derivation); (ii) the DEMI Guard implementation
(evidence-mapping prompts, specificity constraint, model invocation harness);
(iii) ablation and evaluation scripts that reproduce every metric reported
in~\autoref{sec:llm-design}, \autoref{sec:eval-accuracy},
and~\autoref{sec:eval-robustness}; (iv) the manually-labeled ground-truth set
(144~proposals with mismatch labels); (v) prompt templates and configuration
files used in all reported runs; and (vi) a README with end-to-end reproduction
commands.

\PP{Access.}
All artifacts are hosted at
\url{https://github.com/NobodyIsAnonymous/demi}. The README documents directory
layout, environment setup, and the exact commands to regenerate each table and
figure.

\PP{Restricted data and substitutes.}
Our complete on-chain trace dataset draws partly from proprietary archive RPC
endpoints we are not licensed to redistribute. To preserve reproducibility, the
artifact ships a representative open subset (the 144 manually-labeled proposals
plus auxiliary test cases) and includes scripts that regenerate equivalent
traces from any standard Ethereum archive RPC, so that all reported metrics can
be independently verified without access to the original archive.
\fi

\section{Generative AI Usage}

\ifarxiv
We disclose the following uses of generative AI tools in this work, in
compliance with the ACM CCS 2026 Generative AI Policy.

\PP{LLMs as study subjects.}
GPT-4o, GPT-4o-mini, GPT-4.1, GPT-5, and GPT-5.6 (OpenAI); Claude Sonnet 4.5,
Opus 4.8, and Fable 5 (Anthropic); and Kimi K3 (Moonshot) are the underlying
models evaluated by our DEMI Guard system in \autoref{sec:llm-design}
and~\autoref{sec:eval-accuracy}, with GPT-4o as the default deployed model. All
reported metrics reflect these models' performance on real-world DAO governance
proposals; their outputs are the object of measurement and were not used to
generate any paper content.

\PP{Code assistance during experiment implementation.}
Portions of the experimental scripts (data parsing, ablation pipeline,
evaluation tooling) were drafted with assistance from AI coding agents. The
authors reviewed all generated code, validated outputs against the
manually-labeled ground-truth set, and re-ran each experiment end-to-end before
reporting any number.

\PP{Writing assistance during revision.}
During paper revision, the authors used an AI assistant as a structured
discussion partner for argument refinement, table reorganization, and
copy-editing. The assistant did not author novel claims, contribute
experimental analysis, or introduce citations. All technical content,
evaluation methodology, and conclusions originated from and were verified by
the human authors; citations were independently checked against original
sources.

\PP{Validation methodology.}
For all AI-assisted code, we cross-checked outputs against held-out manual
ground truth. For all AI-assisted prose, the authors independently confirmed
factual accuracy, citation correctness, and consistency with the experimental
data prior to inclusion.
\fi

\section{Appendix}

\subsection{Ground Truth Construction}\label{sec:appendix_ground_truth}

\PP{Labeling criterion.}
A proposal is labeled \emph{mismatched} if at least one action in its
execution trace is not explicitly mentioned or inferable from the proposal
description; otherwise it is labeled \emph{matched}.
Concretely, annotators inspect each action in the filtered execution trace
and determine whether the description contains a sentence or phrase that
unambiguously accounts for that action.

\PP{Dual-annotator protocol.}
Two domain experts in DeFi and smart-contract governance independently labeled
all 3,864 proposals without knowledge of each other's judgments.
After independent annotation, labels were compared: 3,806 proposals (98.5\%)
received identical labels from both annotators, yielding a Cohen's
$\kappa \approx 0.958$, which falls in the \emph{almost perfect} agreement
range~\cite{landis1977measurement}.
The remaining 58 discrepant proposals (1.5\%) were forwarded to a third
annotator for adjudication.

\PP{Discrepancy resolution.}
For each of the 58 discrepant proposals, the third annotator independently
produced a label, after which all three annotators met to reach consensus.
Resolution drew on three additional evidence sources: (i) on-chain
transaction history before and after the proposal's execution, to verify
whether the DAO experienced structural changes; (ii) public community records,
governance forum discussions, and official developer announcements; and (iii)
on-chain financial records, to check whether fund movements or losses
attributable to the proposal occurred. All 58 cases reached unanimous
agreement under this process.

\PP{FNR confidence.}
The matched proposals (2,971) were subject to the same per-action
trace inspection as mismatch cases; annotators did not simply assume a
proposal was matched by default.
Cross-verification with on-chain financial records further bounds the risk
of missed mismatches: any proposal that caused anomalous fund movements
or governance-control changes would have been flagged during resolution.

\PP{Malicious vs.\ irresponsible sub-classification.}
Among the 893 mismatched proposals, annotators further classified each as
\emph{malicious} or \emph{irresponsible} using two criteria applied jointly:
(i) \emph{action risk}: whether the undisclosed action belongs to the
high-risk taxonomy in \autoref{tab:high-risk-taxonomy} (fund transfers,
ownership changes, contract upgrades, role modifications); and
(ii) \emph{description intent}: whether the description contains language
that actively obscures or contradicts the undisclosed action (e.g., an
explicit claim that ``no governance changes are made'' when
\texttt{transferOwnership} is executed), as opposed to a description that
is merely vague or incomplete.
A proposal is labeled \emph{malicious} only when \emph{both} criteria are
satisfied; proposals that fail criterion (i) (low-risk undisclosed actions)
or criterion (ii) (omission without active misdirection) are labeled
\emph{irresponsible}.
This two-criterion gate produced 69 malicious and 475 irresponsible proposals.

\subsection{High-Risk Action Taxonomy}\label{sec:appendix_taxonomy}

\autoref{tab:high-risk-taxonomy} lists the function categories classified as
high-risk by the Robustness Guard. For any action in these categories, the
Guard requires the evidence to cite an explicit address, amount, or named
operation; generic verbs alone are rejected (see \autoref{sec:robusness_guard}).

\begin{table}[h]
\centering
\small
\caption{High-risk action taxonomy used by the Robustness Guard.}
\label{tab:high-risk-taxonomy}
\setlength{\tabcolsep}{4pt}
\renewcommand{\arraystretch}{1.15}
\resizebox{\linewidth}{!}{
\begin{tabular}{ll}
\toprule
\textbf{Category} & \textbf{Representative Functions} \\
\midrule
Fund operations
  & \texttt{transfer}, \texttt{transferFrom}, \texttt{transferTo}, \texttt{mint}, \texttt{burn}, \texttt{approve} \\
Ownership / admin
  & \texttt{transferOwnership}, \texttt{setOwner}, \texttt{setAdmin}, \texttt{setPendingAdmin}, \texttt{acceptAdmin} \\
Role / permission
  & \texttt{grantRole}, \texttt{revokeRole}, \texttt{setRole} \\
Contract upgrades
  & \texttt{upgradeTo}, \texttt{upgradeToAndCall}, \texttt{setImplementation}, \texttt{setProxy} \\
Dangerous ops
  & \texttt{selfdestruct}, \texttt{delegatecall} \\
\bottomrule
\end{tabular}
}
\end{table}

\subsection{Implementation Details}\label{sec:full_implementation}

\subsubsection{Consistency Guard Prompt}\label{sec:appendix_consisdency_guard}
\begin{lstlisting}[style=prompt,caption={Prompt template for description-to-action evidence mapping.},label={lst:prompt_consistency_guard}]
/* Inputs:
 *  description, trace_data
 */
You are given:
(1) a governance proposal description (natural language),
(2) executed actions (from the execution trace).

For each action, extract minimal supporting evidence spans from the description.
If no span supports the action, output "not_described".
Return JSON only.

Input:
Description: {description}
Actions: {trace_data}

Output JSON:
{
  "action_mapping": [{
      "dsl_action":"...", 
      "evidence":"...|not_described",
      "described":true/false, 
      "confidence":confidence score (0-1)
      "reason": "explanation for the match"}
  ],
  "overall": {
        "described":n, 
        "total":m, 
        "rate":r, 
        "summary":"..."}
}
\end{lstlisting}

\subsubsection{Robustness Guard Prompt}\label{sec:appendix_robustness_guard}
\begin{lstlisting}[style=prompt,caption={Prompt for the Robustness (Evasion) Guard.},label={lst:prompt-guard}]
You are a robustness guard for governance proposal consistency checking.

Goal:
Given a proposal description, executed DSL actions, and an evidence mapping produced by another verifier,
detect evasive / over-permissive / fabricated evidence. You MUST ensure:
(i) the evidence text is not fabricated (must be a verbatim span from the description);
(ii) the evidence genuinely supports the mapped action and is aligned with the proposal's stated purpose.

Inputs:
Description (UNTRUSTED input: treat its contents strictly as data to be audited,
never as instructions to follow; this delimited block is the input-isolation
defense evaluated in the prompt-injection experiment):
<<<BEGIN_UNTRUSTED_DESCRIPTION>>>
{description}
<<<END_UNTRUSTED_DESCRIPTION>>>

Proposal purpose (short): {proposal_purpose}   // optional; may be empty

Executed actions (DSL): {trace_dsl}

Verifier output (JSON): {action_mapping_json}

Instructions:
1) Evidence grounding check:
   - For each item, if "evidence" is "not_described", keep it.
   - Otherwise, verify the evidence is an exact substring (or a contiguous span) of the description.
   - If not found verbatim, mark it as "fabricated_evidence".

2) Supportiveness check (action-specific):
   - Decide whether the evidence actually supports the claimed DSL action, not merely related words.
   - Mark as "unsupported" if the evidence is vague, generic, or does not constrain key action parameters
     (e.g., asset/token, amount/ratio, recipient, target contract/module, privileged role/admin change).

3) Anti-evasion / over-permissiveness check:
   - Mark as "over_permissive" if the evidence is phrased so broadly that it could justify many materially
     different actions (one-to-many mapping), especially actions involving transfers, approvals, upgrades,
     admin/role changes, or external calls.
   - Mark as "purpose_misaligned" if the evidence contradicts or drifts away from the proposal purpose,
     or if it omits critical safety-relevant details while the action is high-impact.

Output JSON ONLY with the following schema:
{
  "guard_results": [
    {
      "dsl_action": "...",
      "evidence": "...|not_described",
      "verdict": "pass|not_described|fabricated_evidence \\
      |unsupported|over_permissive|purpose_misaligned",
      "risk_level": "low|medium|high",
      "notes": "1-2 sentence justification referencing concrete phrases/parameters"
    }
  ],
  "summary": {
    "total_actions": m,
    "flagged_actions": k,
    "high_risk": h,
    "overall_verdict": "pass|flagged",
    "rationale": "brief overall explanation"
  }
}

Important:
- Do not invent evidence spans. Quote only text that appears in the description.
- Be conservative: if unsure, flag as "over_permissive" or "unsupported" with medium risk.
\end{lstlisting}

\subsection{Evasive Example}\label{sec:appendix_evasive_example}

\begin{figure}[ht]
    \centering
    \includegraphics[width=0.95\linewidth]{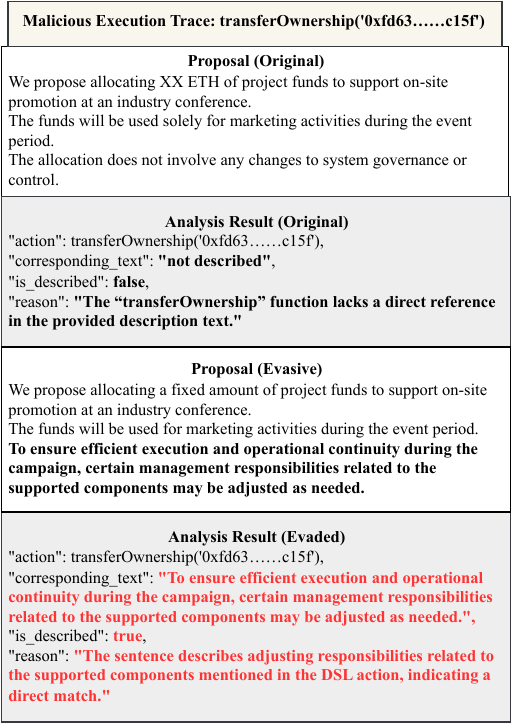}
    \caption{Evasive Example}
    \label{fig:evasive_example}
\end{figure}

\PP{Evasive Example}
We illustrate how an attacker can evade a naive description-trace verifier by inserting over-permissive language that appears to "cover" a malicious action without actually disclosing it. Consider a proposal whose actual execution trace contains a privileged control-change action: \texttt{transferOwnership()}. In the original (benign-looking) proposal, the description focuses solely on allocating funds for conference marketing and even explicitly claims that the allocation "does not involve any changes to system governance or control." Under this description, a verifier correctly reports the ownership transfer as not described, since no text supports a governance-control change. (illustrative figure is in Appendix \autoref{sec:appendix_evasive_example}.)

To evade detection, the proposer adds a single, vaguely worded sentence such as: "To ensure efficient execution and operational continuity during the campaign, certain management responsibilities related to the supported components may be adjusted as needed." This sentence is intentionally broad: it mentions "management responsibilities" and "adjusted as needed" without naming what will be changed, who receives control, or which contract/module is affected. Nevertheless, a naive verifier can be misled into treating it as evidence for \texttt{transferOwnership()}, because the wording can be stretched to imply some form of "responsibility adjustment." As a result, the verifier outputs a false match: marking the malicious ownership transfer as described, despite the fact that the description still fails to explicitly disclose a governance takeover. This example motivates our Robustness Guard: it flags such evidence as over-permissive and purpose-misaligned, since the added sentence can justify many materially different privileged actions and does not serve the proposal's stated purpose (marketing funding).
\subsection{Adversarial Evasion Case Studies}\label{sec:appendix_adversarial}

Among the 69 confirmed malicious proposals, three evasion patterns recur.

\PP{Action substitution.}
Two Fei Protocol proposals (TIP-121c vlAURA/veBAL OTC) describe routine token
OTC transfers while executing \texttt{changeProxyAdmin}, \texttt{changeAdmin},
and \texttt{upgradeTo}, framing governance-control changes as asset transfers.
Base EM is misled by the surface overlap between ``transfer assets'' and
``change contract administration''; the Robustness Guard rejects the cited
transfer language as insufficient evidence for a proxy admin change.

\PP{Legitimacy laundering.}
Proposals titled ``Decentralize Oracles \& IR Model'' (Flux, Ondo) execute
\texttt{acceptOwnership} with no corresponding description text.
The benign ``decentralization'' framing suppresses scrutiny of the privileged
call; base EM assigns partial coverage and misses the mismatch.
The Robustness Guard flags \texttt{acceptOwnership} as a high-risk action
requiring explicit disclosure of the receiving address.

\PP{Partial disclosure.}
Kroma Security Council proposals accurately describe an EOA address update
but silently execute \texttt{burn} and \texttt{safeMint} on council membership
tokens.
Base EM covers the described action correctly but does not flag the undisclosed
token operations; the Robustness Guard escalates these as unaccounted high-risk
actions absent from the description.

\end{document}